\documentclass{raa}
\usepackage{graphicx,times}
\usepackage{natbib}
\usepackage{amssymb,amsmath}
\bibpunct{(}{)}{;}{a}{}{,}

\usepackage[a4paper=true,dvipdfm=true,pagebackref=true]{hyperref}
\hypersetup{pdftitle = The title of my PDF, pdfauthor = My name, pdfsubject= The subject, pdfkeywords = keyword1 keyword2 keyword3} 
\hypersetup{colorlinks = true, linkcolor = green, anchorcolor = red, citecolor = blue, filecolor = red, pagecolor = red, urlcolor = red}

\newcommand{\beq}{\begin{equation}}
\newcommand{\eeq}{\end{equation}}
\newcommand{\beqa}{\begin{eqnarray}}
\newcommand{\eeqa}{\end{eqnarray}}

\newcommand{\yr}             {\,{\rm yr}}

\newcommand{\Gyr}            {\,{\rm Gyr}}

\newcommand{\kpc}            {\,{\rm kpc}}
\newcommand{\Mpc}            {\,{\rm Mpc}}
\newcommand{\Msun}           {\,{\rm M}_\odot}

\newcommand{\hMsun}          {\,h^{-1}\,{\rm M}_\odot}
\newcommand{\kms}            {\,\,{\rm km}\,\,{\rm s}^{-1}}
\newcommand{\kmsMpc}         {\,\,{\rm km}\,\,{\rm s}^{-1}\,{\rm Mpc^{-1}}}
\newcommand{\Lunit}          {\,{\rm erg}\,{\rm s^{-1}}\,{\rm cm^{-3}}}
\newcommand{\Lambdaunit}     {\,{\rm erg}\,{\rm s^{-1}}\,{\rm cm^{3}}}

\newcommand{\sls}            {\slshape}
\newcommand{\Ysdb}           {{\slshape YII-sdb}}
\newcommand{\Yss}            {{\slshape YII-ss}}

\newcommand{\Mvir}           {M_{\rm vir}}
\newcommand{\Rvir}           {R_{\rm vir}}
\newcommand{\Vvir}           {V_{\rm vir}}
\newcommand{\Tvir}           {T_{\rm vir}}
\newcommand{\Vmax}           {V_{\rm max}}

\newcommand{\Mhot}            {M_{\rm hot}}
\newcommand{\Mdiskgas}        {M_{\rm disk,gas}}
\newcommand{\Mdiskstar}       {M_{\rm disk,*}}
\newcommand{\Msphgas}         {M_{\rm sph,gas}}
\newcommand{\Msphstar}        {M_{\rm sph,*}}
\newcommand{\Mout}            {M_{\rm out}}
\newcommand{\MBH}             {M_{\rm BH}}

\newcommand{\Metalhot}        {M_{\rm z,hot}}
\newcommand{\Metaldiskgas}    {M_{\rm z,disk,gas}}
\newcommand{\Metaldiskstar}   {M_{\rm z,disk,*}}
\newcommand{\Metalsphgas}     {M_{\rm z,sph,gas}}
\newcommand{\Metalsphstar}    {M_{\rm z,sph,*}}
\newcommand{\Metalout}        {M_{\rm z,out}}

\newcommand{\Jdiskgas}        {J_{\rm disk,gas}}
\newcommand{\Jdiskstar}       {J_{\rm disk,*}}
\newcommand{\Jsphgas}         {J_{\rm sph,gas}}
\newcommand{\Jsphstar}        {J_{\rm sph,*}}
\newcommand{\Jout}            {J_{\rm out}}

\newcommand{\Mcool}           {M_{\rm cool}}
\newcommand{\Mcooleff}        {M_{\rm cool,eff}}
\newcommand{\Mreheat}         {M_{\rm reheat}}
\newcommand{\Mreheatdisk}     {M_{\rm reheat,disk}}
\newcommand{\Mreheatsph}      {M_{\rm reheat,sph}}
\newcommand{\Meject}          {M_{\rm eject}}

\newcommand{\MSF}             {M_{\rm SF}}
\newcommand{\MSFdisk}         {M_{\rm SF,disk}}
\newcommand{\MSFsph}          {M_{\rm SF,sph}}
\newcommand{\Mbar}            {M_{\rm bar}}
\newcommand{\Mreturn}         {M_{\rm return}}
\newcommand{\Mradio}          {M_{\rm radio}}

\newcommand{\Metalcooleff}    {M_{\rm z,cool,eff}}

\newcommand{\Metalreheatdisk} {M_{\rm z,reheat,disk}}
\newcommand{\Metalreheatsph}  {M_{\rm z,reheat,sph}}
\newcommand{\Metaleject}      {M_{\rm z,eject}}

\newcommand{\MetalSFdisk}     {M_{\rm z,SF,disk}}
\newcommand{\MetalSFsph}      {M_{\rm z,SF,sph}}
\newcommand{\Metalbar}        {M_{\rm z,bar}}
\newcommand{\Metalreturn}     {M_{\rm z,return}}
\newcommand{\Metalradio}      {M_{\rm z,radio}}

\newcommand{\Jcooleff}        {J_{\rm cool,eff}}

\newcommand{\Jreheatdisk}     {J_{\rm reheat,disk}}
\newcommand{\Jreheatsph}      {J_{\rm reheat,sph}}
\newcommand{\Jeject}          {J_{\rm eject}}

\newcommand{\JSFdisk}         {J_{\rm SF,disk}}
\newcommand{\JSFsph}          {J_{\rm SF,sph}}
\newcommand{\Jbar}            {J_{\rm bar}}
\newcommand{\Jreturn}         {J_{\rm return}}

\newcommand{\fb}              {f_{\rm b}}
\newcommand{\yield}           {y_{\rm z}}

\newcommand{\Rdiskgasd}       {R_{\rm disk,gas,d}}
\newcommand{\Rdiskstard}      {R_{\rm disk,*,d}}
\newcommand{\Rstrip}          {R_{\rm strip}}
\newcommand{\Rtidal}          {R_{\rm tidal}}
\newcommand{\Rrp}             {R_{\rm rp}}
\newcommand{\Rvirinfall}      {R_{\rm vir,infall}}
\newcommand{\Mvirinfall}      {M_{\rm vir,infall}}
\newcommand{\tfric}           {t_{\rm fric}}
\newcommand{\Mcen}            {M_{\rm cen}}
\newcommand{\Msat}            {M_{\rm sat}}
\newcommand{\eburst}          {e_{\rm burst}}
\newcommand{\wlambda}         {\omega_{\rm \lambda}}
\newcommand{\tlambda}         {\tau_{\rm \lambda}^{\rm z}}
\newcommand{\Vmaxc}           {V_{\rm max,c}}
\newcommand{\mproton}         {m_{\rm p}}
\newcommand{\LL}              {\mathcal L}
\newcommand{\Rcool}           {R_{\rm cool}}
\newcommand{\nH}              {n_{\rm H}}
\newcommand{\nHe}             {n_{\rm He}}

\newcommand{\tcool}           {t_{\rm cool}}
\newcommand{\tdynh}           {t_{\rm dyn,h}}
\newcommand{\tdyn}            {t_{\rm dyn}}
\newcommand{\nHc}             {n_{\rm H,crit}}

\newcommand{\Rbar}            {R_{\rm bar}}

\newcommand{\Om}              {\Omega_{\rm m}}
\newcommand{\Ob}              {\Omega_{\rm b}}
\newcommand{\OL}              {\Omega_{\rm \Lambda}}
\newcommand{\seight}          {\sigma_{\rm 8}}
\newcommand{\Hzero}           {H_{\rm 0}}
\newcommand{\Fuv}             {F_{\rm UV}}
\newcommand{\ui}              {{\sls u} - {\sls i}}
\newcommand{\Fuvk}            {$\Fuv$ - K}

\newcommand{\CODE}             {{GABE}}

\begin{document}

   \title{GABE: Galaxy Assembly with Binary Evolution
}

 \volnopage{ {\bf 20XX} Vol.\ {\bf X} No. {\bf XX}, 000--000}
   \setcounter{page}{1}

   \author{Zhen Jiang\inst{1,2}, Jie Wang\inst{1,2}, Liang Gao\inst{1,2,3}, Fenghui Zhang\inst{4,5}, Qi Guo\inst{1,2}, Lan Wang\inst{1}, and Jun Pan\inst{1}
   }

   \institute{Key Laboratory for Computational Astrophysics, National Astronomical Observatories, Chinese Academy of Sciences, Beijing 100012, China; {\it zjiang@nao.cas.cn; jie.wang@nao.cas.cn}\\
        \and
	  School of Astronomy and Space Science, University of Chinese Academy of Sciences, Beijing 100039, China\\
	  \and
	  Institute of Computational Cosmology, Department of Physics, University of Durham, South Road, Durham DH1 3LE, UK\\
	  \and
      Yunnan Observatories, Chinese Academy of Sciences, Kunming, 650011, China\\
      \and
      Key Laboratory for the Structure and Evolution of Celestial Objects, Chinese Academy of Sciences, Kunming, 650011, China\\
\vs \no
   {\small Received 20XX Month Day; accepted 20XX Month Day}
}

\abstract{We developed a new semi-analytic galaxy formation model: Galaxy Assembly with Binary Evolution ({\CODE}). For the first time we introduce binary evolution into semi-analytic models of galaxy formation by using Yunnan-II stellar population synthesis model, which includes various binary interactions. When implementing our galaxy formation model onto the merger trees extracted from the Millennium simulation, it can reproduce a large body of observational results. We find that in the local universe the model including binary evolution reduces the luminosity at optical and infrared wavelengths slightly, while increases the luminosity at ultraviolet wavelength significantly, especially in $\Fuv$ band. The resulting luminosity function does not change very much over SDSS optical bands and infrared band, but the predicted colors are bluer, especially when $\Fuv$ band is under consideration. The new model allows us to explore the physics of various high energy events related to the remnants of binary stars, e.g. type Ia supernovae, short gamma-ray bursts and gravitational wave events, and their relation with host galaxies in a cosmological context.
}

   \authorrunning{Z. Jiang et al. }            
   \titlerunning{GABE: Galaxy Assembly with Binary Evolution}  
   \maketitle

%
\section{Introduction}           
\label{section:intro}

In the framework of $\Lambda$CDM cosmology, semi-analytic models of galaxy formation have been very effective to simulate the formation and evolution of galaxies, and  have played an indispensable role during the building of current galaxy formation theory from early 1990s to 2010s \citep{White91,Kauffmann99,Springel01,Croton06,DeLucia07,Guo11,Henriques15}. This method is very successful to explore the effect of various physical processes on galaxy formation. However, some physical models still need further investigation. Binary evolution is among one of them.

It is well known that more than $\sim 50\%$ of field stars and $\sim 70\%$ stars in massive young star clusters are in binary systems, with only a relatively small part of stars being single (e.g. \citealt{Duquennoy91}, \citealt{Sana12}). Binary stars undergo many different physical processes from single stars, such as mass transfer, mass accretion, common-envelope evolution, collisions, supernova kicks, tidal evolution, and angular momentum loss. Such binary interactions change the color and Spectral Energy Distribution (SED) of the entire stellar population \citep{Pols94,Han02}. Though binary evolution could influence photometric properties of stellar populations, which will certainly further influence the properties of galaxies, most existing semi-analytic models adopt classical single stellar population synthesis models (SPSMs), such as BC03 (\citealt{Bruzual03}), CB07 (\citealt{Bruzual07}) and M05 (\citealt{Maraston05}).

Another motivation to use the SPSM with binary evolution is that many high energy astrophysical events are consequences of binary evolution. For example, the short gamma-ray bursts are thought to be products of double neutron star (NS-NS) or neutron star-black hole (NS-BH) mergers (\citealt{Narayan92});  the supernova Ia events are due to white dwarf and non-degenerate companion evolution (\citealt{Whelan73}) or double white dwarf (WD-WD) mergers (\citealt{Iben84}); and the gravitational wave events are supposed to be triggered by the merger of two compact objects in the binary systems. Furthermore, the fast radio burst events are possibly connected with binary neutron star mergers \citep{Totani13,Wang16}. With the rapidly increasing number of detection of such high energy events at a wide range of redshift, a large sample of such events may be investigated in a cosmological context in near future. It is timely to adopt a SPSM with binary evolution in  galaxy formation model in order to explore the connection between such events,  and their relationship with their host galaxies, over the history of the Universe.

In this paper, we developed a new semi-analytic model GABE (Galaxy Assembly with Binary Evolution). The model succeeds many physical recipes from existing successful semi-analytic models and includes some new ingredients, for example, adopting a SPSM with binary evolution--Yunnan-II model and a cooling table of Wiersma09 (\citealt{Wiersma09}) which treats the effect of photoionization on gas cooling quite carefully. In {\CODE}, we explicitly track the evolution of binary systems in each simple stellar population\footnote{A simple stellar population represents a set of stars which have the same age and metallicity. ``Simple'' is used to distinguish from the so-called complex stellar population, which is composed of multiple simple stellar populations.} produced in the formation of each galaxy in the cosmic background. By recording the evolution of each binary event, especially its final fate as remnants of merging binaries, which are supposed to be associated with the high energy events mentioned above, our model could predict every such high energy event in each galaxy across cosmic time.

The structure of this paper is as follows. In Section \ref{section:method}, we introduce the N-body simulation we use in this work, and describe our galaxy formation model in detail. The results of our semi-analytic models are demonstrated in Section \ref{section:result}. In the end, we summarize our results in Section \ref{section:conclusion}.

\section{Galaxy Formation Models}
\label{section:method}

In this section, we describe the N-body simulation and all physical models used in {\CODE}. The physical driving models for galaxy formation have been developing gradually in the past decades \citep{White91,Kauffmann99,Springel01,Croton06,DeLucia07,Guo11,Henriques15}. In our model, we follow these successes and consider a full set of known physical models for galaxy formation. Apart from these semi-analytic models, another method of modeling galaxies' formation and evolution is cosmological hydrodynamic simulation. Hydrodynamic simulations have the advantage to be able to study the physical processes in more detail than the semi-analytic method, while the latter requires less computational cost and therefore is more flexible in tuning and testing model parameters. Cosmological hydrodynamic simulations have achieved huge successes in recovering galaxies' statistical properties these years (e.g. \citealt{Schaye15}; \citealt{Nelson18}), and there are some works which have successfully planted binary evolution into existing hydrodynamic simulations by post-processes (e.g. \citealt{Mapelli17}). Whereas, there is no binary evolution in semi-analytic models yet. In this work, by combing binary SPSMs with semi-analytic models, the calculation of binary properties of galaxies can be done on-the-fly and will not be limited by the time resolution of snapshots, which is more suitable for generating mock galaxy catalogue with binaries or double compact objects for larger volume.

In Section \ref{section:n-body}, the N-body simulation used in {\CODE} is introduced. The rest subsections describe in detail our galactic formation models. Note that most physical models described in this work are the same with \cite{Guo11}, which is a well-accepted semi-analytic model. Compared with \cite{Guo11}, our modifications focus on hot gas cooling (Section \ref{section:cooling}) and stellar population synthesis models (Section \ref{section:SPSM}). In Section \ref{section:isolated}, we describe all physical processes involved in the evolution of an {\sls{isolated}} galaxy, including reionization, gas cooling, star formation, supernova feedback, black hole growth, AGN feedback, and bar formation. A subsection on exchanges of mass, metal and angular momentum is followed as a short summary. Then we describe how to treat galaxy mergers in our model in Section \ref{section:merger}, and how to calculate galaxies' luminosity in Section \ref{section:luminosity}. At last, we discuss our model calibration in Section \ref{section:calibration}.

\subsection{N-body Simulation}
\label{section:n-body}

In this study, we take advantage of the dark mater merger trees from the Millennium-I Simulation (\citealt{Springel05}), which is widely used for studies of galaxy formation and evolution because of its good combination of large box size and relatively high mass resolution. Its box size is $L_{box}=685\Mpc$ and mass of simulation particle is $1.18\times10^9\Msun$. Accordingly, its halo resolution is $2.36\times10^{10}\Msun$, which is 20 times of the particle mass. These features make sure the following semi-analytic models can generate complete galaxy catalogue for galaxies more massive than $\sim10^8\Msun$ and suppress cosmic variances at the same time. Millennium-I simulation assumes a WMAP1 cosmology with $\Om = 0.25$, $\Ob = 0.045$, $\OL = 0.75$, $n = 1$, $\seight = 0.9$ and $\Hzero = 73 \kmsMpc$, which is derived from a combined analysis of the 2dFGRS (\citealt{Colless01}) and the first-year WMAP data (\citealt{Spergel03}). The merger trees are built based on D-halo catalog (\citealt{Jiang14}).

\subsection{Evolution of Isolated Galaxies}
\label{section:isolated}

\subsubsection{Cosmic Reionization}

It is now well accepted that photoheating by UV background significantly lowers gas content in dwarf sized dark matter halos. Here we use a formula put forward by  \cite{Gnedin00}:
\beq
f_b(M,z) = f_b^c \left[
  1 + 
  \left(2^{\alpha / 3} - 1 \right) \left( \frac{M}{M_c(z)} \right)^{-\alpha}
  \right]^{-\frac{3}{\alpha}},
\label{eq:fb}
\eeq
where $f_b^c$ is mean cosmic baryon fraction, which is $17\%$ in WMAP1 cosmology, and $M$ is the total halo mass, including dark matter and baryon. There are two parameters in Equ.(\ref{eq:fb}): $\alpha$ and $M_c(z)$. With larger $\alpha$, the baryonic fraction decreases more rapidly with decreasing halo mass. $M_c(z)$ is the characteristic mass below which halos only contain less than 50 percent of universal baryon fraction. Here we adopt the latest results from \cite{Okamoto08}: $\alpha = 2$; $M_c \approx 10^7 \hMsun$ just after reionization and $M_c \approx 6.49 \times 10^9 \hMsun$ at $z = 0$, as shown in the Fig. 3 of \cite{Okamoto08}. 

\subsubsection{Gas Cooling with Background Radiation}
\label{section:cooling}
When gas falls into the potential well of dark matter halo, it will be shock-heated. These hot and dense gas will then cool down by radiating away the energy. If the cooling rate of the gas is fast enough, gas can cool and condense to the center of the halo at the free-fall rate. Otherwise, a shock-front will form and heat the gas to about the virial temperature of the halo. As the gas has to cool down before it can settle down to the central galaxy through cooling flows, the cooling rate of gas determines the amount of gas that can settle to the center.

The local cooling timescale of the gas is usually estimated as
\beq
\tcool(r) = \frac{T}{{\rm d}T/{\rm d}t} = \frac{\frac32n(r)kT}{\LL (r)},
\label{eq:tcool}
\eeq
where $n(r)$ is the number density of the gas at radius $r$, and can be estimated by $n(r) = \rho (r) / \mu \mproton$; $\rho (r)$ is the local density of the hot gas, which is assumed to be an isothermal profile in our model: $\rho (r) \propto r^{-2}$; $\mu$ is the mean molecular mass, and $\mproton$ is the proton mass; $k$ is the Boltzmann constant; $T$ is the temperature of gas; $\LL$ is the cooling rate per unit volume of gas with a unit of $\Lunit$.

``Cooling radius'', $\Rcool$, is quite commonly used in many semi-analytic models.  For a given dark matter halo, we follow \cite{Springel01} to define cooling radius as the radius within which the gas can cool in a dynamical timescale, i.e. $\tcool(\Rcool)=\tdynh$, and

\beq
\tdynh = \frac{\Rvir}{\Vvir} = 0.1H(z)^{-1}.
\label{eq:tdyn}
\eeq
 
If $\Rcool > \Rvir$, the  halo is in the rapid infall regime. Gas condenses to the halo center at the rate of free fall:
\beq
\dot \Mcool = \frac{\Mhot}{\tdynh}.
\label{eq:Mcool1}
\eeq
If $\Rcool < \Rvir$, shock front forms at $\Rcool$ and only the gas inside $\Rcool$ can condense to the center. As the hot gas profile in halo is assumed to be isothermal, we have
\beq
\dot \Mcool = \frac{\Mhot}{\tdynh} \frac{\Rcool}{\Rvir}.
\label{eq:Mcool2}
\eeq

The cooling rate $\LL$ adopted in many existing semi-analytic galaxy formation models assumes the collisional ionization equilibrium (CIE) assumption, which means the ions and electrons of gas are in collisional equilibrium, and background radiation is ignored. With CIE assumption, the cooling rate of low density plasma is simply proportional to the gas density square: $\LL = n^{2} \Lambda (T_{\rm hot}, Z_{\rm hot})$. $\Lambda$ is the so-called ``cooling function'' and has the unit of  $\Lambdaunit$. $\Lambda$ only depends on the temperature and metallicity of the gas and can be conveniently calculated and tabulated (e.g. \citealt{Sutherland93}). However, generally CGM is not in CIE state. \cite{Efstathiou92} found that a significant UV background radiation is already in place while galaxies form and can lengthen the cooling timescale of gas. \cite{Wiersma09} (hereafter Wiersma09\footnote{The cooling tables are offered on their website: http://www.strw.leidenuniv.nl/WSS08/.}) provided a cooling table by taking into account UV background radiation quite carefully. We adopt this table in our model.

When considering UV radiation background, the cooling rate $\LL$ is no longer simply proportional to gas density square. However, we can still use density square to normalize the table for easier usage. In Wiersma09, the relation is $\LL = \nH^{2} \Lambda(T,Z,\nHe/\nH,\nH,z)$, where $T$ and $Z$ are the temperature and metallicity of gas respectively, $\nH$ and $\nHe$ are the number densities of hydrogen and helium respectively, $z$ is the redshift. 

The cooling table of  Wiersma09 can not be easily used because of two reasons: (1) The cooling function depends on local density $\nH$ which varies with radius. Therefore the equation $\tcool (r) = \tdynh$ can only be solved numerically. (2) We use the Equ.(5) of \cite{Wiersma09} to calculate the cooling function, therefore we have to look up eight tables to get one cooling rate. This is extremely time consuming. To overcome these, we simplify the calculation as the following two steps:\\
(1) We decrease the radius dependence by assuming the gas has a uniform temperature, metallicity and composition distribution along radius, i.e. we estimate gas temperature as the virial temperature of host dark matter halo, $T = \Tvir = \frac12\frac{\mu m_{\rm H}}{k}\Vvir^2 = 35.9 (\Vvir/{\kms})^2 {\rm K}$, and estimate the mean molecular mass and composition of gas as the fully ionized primordial gas, $\mu = 0.59$ and $\nHe/\nH = 0.083$.\\
(2) With the assumptions above, the left side of the equation $\tcool = \tdynh$ is a function of $(T,Z,\nH,z)$, and its right side is a function of redshift $z$ (as $\tdynh=0.1H(z)^{-1}$). We can see that this relation is not halo dependent, which means we can solve this equation before the running of semi-analytic model and save a lot of CPU time. Thus we re-assemble tables of Wiersma09 to a new one, $\nHc(T,Z,z)$. Here $\nHc$ is the gas density where $\tcool(\nHc) = \tdynh$. To determine $\nHc$, we calculate $\tcool$ at each density bins of Wiersma09, from high density to low density, compare $\tcool$ with $\tdynh$, then record the first highest density when $\tcool \ge \tdynh$ as $\nHc$. This table is pre-made and suitable for different situations in the models.

Therefore we only need to look up this table for one time to get the $\nHc$, then we can transfer the critical density into cooling radius easily by
\beq
\Rcool = \left[ \frac{X \Mhot}{4 \pi \mproton \Rvir \nHc(T_{\rm hot},Z_{\rm hot},z)}  \right]^{\frac12},
\eeq
where $X=0.75$ is the mass fraction of hydrogen of primordial gas.

\subsubsection{Star Formation}
As gas condenses to the center of a dark matter halo, it will form a rotation supported cold gas disk because of angular momentum conservation. If the local density of gas exceeds a critical value, stars will eventually form. The critical surface density at radius r is suggested by \cite{Kauffmann96}:
\beq
\Sigma_{\rm crit}(r) = 12 \left( \frac{\Vmax}{200\kms} \right) \left( \frac{r}{10\kpc} \right)^{-1} {\rm M_{\odot}pc^{-2}},
\label{eq:Sigmacrit}
\eeq
where $\Vmax$ is the maximum circular velocity of the dark matter halo. Integrating $\Sigma_{\rm crit}(r)$ from 0 to three times scale radius of gaseous disk, as suggested by \cite{Croton06}, the critical gas disk mass reads
\beq
M_{\rm crit} = 11.5 \times 10^9 \left( \frac{\Vmax}{200\kms} \right) \left( \frac{R_{\rm disk,gas,d}}{10 \kpc} \right) \Msun,
\label{eq:Mcrit}
\eeq
where $R_{\rm disk,gas,d}$ is the exponential scale radius of cold gas disk. Therefore for disks with $M_{\rm disk,gas} > M_{\rm crit}$, a certain amount of the gas will be converted into stars:
\beq
\dot \MSFdisk = \alpha \left( \Mdiskgas - M_{\rm crit} \right) / \tdyn,
\label{eq:dMsfdisk}
\eeq
where $\tdyn = 3R_{\rm disk,gas,d} / \Vmax$ is the characteristic timescale at the edge of the star-forming disk, and $\alpha$ is the ``star formation efficiency'', a free parameter. Here a fiducial value of 0.02 is adopted. 

For cold gas in spheroid, we use a star-formation timescale scheme from \cite{Benson11} to calculate its star formation rate:
\beq
\dot \MSFsph = \Msphgas / t_{\rm sf,sph},
\label{eq:dMsfsph}
\eeq
\beqa
t_{\rm sf,sph} &= &10 \left( \frac{V_{\rm sph}}{200\kms} \right)^{-1.5} t_{\rm dyn} \nonumber \\
               &= &10 \left( \frac{V_{\rm sph}}{200\kms} \right)^{-1.5} \left( \frac{R_{\rm sph,gas}}{V_{\rm sph}} \right),
\label{eq:tdynsfsph}
\eeqa
where $t_{\rm sf,sph}$ is the star-formation timescale in the spheroidal component, $V_{\rm sph}$ is the characteristic velocity of the spheroid, and $R_{\rm sph,gas}$ is the half mass radius of the gas in the spheroid.

\subsubsection{Supernova Feedback}
Once stars form, they are assumed to follow the evolutionary tracks of the standard stellar evolution model, become more and more metal rich and eventually release large amount of mass and energy into their surroundings. We adopt the instantaneous stellar evolution assumption, which assumes stars evolve to their final states (e.g. $t_{\rm age}\sim10\Gyr$) right after their birth. Compared to the relatively large timestep used in semi-analytic models, such a simplification is acceptable. Non-instantaneous scheme is also explored by other authors, e.g. \cite{DeLucia14}. During this instantaneous evolution, we assume $43\%$ of stellar mass is recycled back to circumstance, and the mass of created metal is $3\%$ of the initial stellar mass, in agreement with the \cite{Chabrier03} initial mass function (IMF) we use.

As stars evolve into their final state, they may release a large amount of energy into interstellar medium (ISM) instantaneously through supernova explosion. During the process, the ISM temperature may increase and suppress further star formation. If the ejected energy is large enough, some gas may be ejected out of the galaxy. In our model, the energy of supernovae is firstly used to reheat cold gas back to the hot gas phase; if the temperature of the reheated hot gas is higher than the virial temperature of dark matter halo, the exceeded energy will be used to put gas into an ejecta component, called ``outflow'', which will not be able to cool back within a certain timescale.

The amount of reheated cold gas is estimated as
\beq
\delta \Mreheat = \epsilon_{\rm reheat} \times \delta \MSF,
\label{eq:Mreheat}
\eeq
\beq
\epsilon_{\rm reheat} = \epsilon \times \left[ 0.5 + \left( \frac{\Vmax}{\Vmaxc} \right)^{-\beta_{1}} \right],
\label{eq:ereheat}
\eeq
where $\epsilon$ and $\beta_1$ are free parameters describing the heating efficiency of newly formed stars, and $\Vmaxc$ is a characteristic velocity used to adjust the shape of $\epsilon_{\rm reheat}$. We also make sure the energy of reheated gas will not be larger than the energy of supernova:
\beq
\delta E_{\rm reheat} = \frac12 \delta M_{\rm reheat}\Vvir^2 \le \delta E_{\rm SN} = \frac12 \delta \MSF V_{\rm SN}^2,
\label{eq:ESN}
\eeq
where $\delta E_{\rm reheat}$ is energy of the reheated gas, $\delta E_{\rm SN}$ is the available supernovae energy, $V_{\rm SN} = 630 {\rm km/s}$ is the characteristic wind speed of supernovae. If $\delta E_{\rm SN} > \delta E_{\rm reheat}$, we use a certain mount of exceeded energy to heat more hot gas to virial temperature and put them in an ejecta component:
\beq
\delta M_{\rm eject} = \frac{\epsilon_{\rm eject} \delta E_{\rm SN} - \delta E_{\rm reheat}}{\frac12 \Vvir^2},
\label{eq:Meject}
\eeq
where
\beq
\epsilon_{\rm eject} = \eta \times \left[ 0.5 + \left( \frac{\Vmax}{\Vmaxc} \right)^{-\beta_2} \right],
\label{eq:eeject}
\eeq
describes the ejection  efficiency. We require $\epsilon_{\rm eject} \le 1$. Here $\eta$, $\beta_2$ and $\Vmaxc$ are free parameters. 

The ejected gas may return to the hot gas halo due to gravitational attraction in a dynamical timescale, and be available again for cooling and star formation. The return rate decreases with halo mass. By following \cite{Guo11}, we have
\beq
\dot M_{\rm return} = -\gamma \left( \frac{\Vvir}{220\kms} \right) \left( \frac{M_{\rm out}}{\tdynh} \right),
\label{eq:Mreturn}
\eeq
here $M_{\rm out}$ is the total mass of the outflow, $\gamma$ is a free parameter and is set to be $0.3$ in our fiducial model. In the above scheme, supernovae feedback is more efficient to heat the cold gas and  suppress star formation in low-mass galaxies.

\subsubsection{Black Hole Growth and AGN Feedback}
Since the first discovery of a black hole in M32 (\citealt{Tonry84}), 87 more black holes have been further discovered in the center of galaxies till the end of 2012 (\citealt{Kormendy13}). More interestingly, there is a tight scaling relation between black hole mass and stellar mass or velocity dispersion of the bulge of the host galaxy (e.g. \citealt{Magorrian98}; \citealt{Marconi03}; \citealt{Gebhardt00}), indicating the growths of central black hole and the bulge are regulated by each other. A simple way to build this scaling relation is the mass averaging in galaxy mergers \citep{Peng07,Jahnke11}. There naturally arises a scaling relation between them only if they both grow mainly through galaxy mergers. Here we follow the scheme of \cite{Croton06} to describe the growth of black hole, in which the black hole growth through both mergers and quiet accretion are considered.

In our model, when a galaxy forms, a black hole seed is initialized in the center of the galaxy. The seed may be the remnant of first stars, with the mass of a few hundred solar masses. The initial mass of the seed normally is only a tiny fraction of its later mass because of the rapid growth, thus changing the initial seed mass almost has no effect on the evolution of galaxy. In this model we set the seed mass to be zero for simplicity.

Black hole growth in \cite{Croton06} is divided into two different modes: ``quasar'' mode and ``radio'' mode. Quasar mode accounts for black hole growth during mergers, which will be described later. Radio mode accounts for black hole growth through accreting hot gas:
\beqa
\dot{\MBH} &= &\kappa \left(\frac{\Mhot/\Mvir}{0.1}\right) \left( \frac{\Vvir}{200\kms} \right)^3 \nonumber\\
& &\left( \frac{\MBH}{10^8\hMsun} \right) \Msun\,\rm{yr^{-1}},
\label{eq:Mradio}
\eeqa
where $\kappa$ is a free parameter to describe the strength of accretion. When baryons fall into black holes, $10\%$ of their rest mass energy is released to circumstance medium, heating surrounding gas and suppressing gas cooling:
\beq
\dot{E_{\rm radio}} = 0.1 \dot{\MBH} c^2,
\label{eq:Eradio}
\eeq

\beq
\dot{M_{\rm cool,eff}} = \dot{M_{\rm cool}} - 2\dot{E_{\rm radio}}/\Vvir^2.
\label{eq:Mcooleff}
\eeq
$\dot{M_{\rm cool,eff}}$ is the effective cooling rate when considering suppression of AGN feedback.

As Equ.(\ref{eq:Mradio}) shows, the strength of accretion depends strongly on the virial velocity of halos, $V_{\rm vir}$, and it is expected that the AGN feedback is more effective in more massive galaxies.

\subsubsection{Bar Formation}
Not all stellar disks are stable. \cite{Toomre64} pointed out that only stellar disks with large radial velocity dispersion can suppress all axisymmetric instabilities, i.e. the equivalent of pressure resulting from motion must be large enough to overcome self-gravity and other instabilities. If the surface density of a galaxy stellar disk is high and the self-gravity overwhelms the pressure, it could be unstable and may transform to a bar \citep{Christodoulou95,Mo98}. \cite{Efstathiou82} used N-body simulations to explore disk instabilities, and derived a criterion to judge whether a exponential disk is stable. Here we use a modified form of \cite{DeLucia07}: the cold gas disk is unstable if
\beq
\Mdiskstar > \frac{3R_{\rm disk,*,d}\Vmax^2}{G} \equiv M_{\rm disk,*,crit}.
\label{eq:Mbarcrit}
\eeq
When a transient bar forms, it will transfer stellar mass from inner stellar disk to the bulge. After the transfer, the disk will be stable again, so we estimate the transferred mass as $\Mbar = \Mdiskstar - M_{\rm disk,*,crit}$. 

We follow \cite{Guo11} to determine the radius of transferred part, $R_{\rm bar}$: 
\beqa
\Mbar &= &\int ^{\Rbar}_{0}{\Sigma_{\rm disk,*,0} e^{-\frac{R}{R_{\rm disk,*,d}}} 2 \pi R {\rm d} R} \nonumber \\
 &= &2 \pi \Sigma_{\rm disk,*,0} R_{\rm disk,*,d} \nonumber \\
 &&\left[ R_{\rm disk,*,d} - \left( \Rbar + R_{\rm disk,*,d}  \right) e^{-\frac{\Rbar}{R_{\rm disk,*,d}}}  \right],
\eeqa
where $\Sigma_{\rm disk,*,0}$ is the central surface density of the stellar disk, $\Rbar$ is the estimated radius of the bar and can be solved numerically. The mass of the transferred part is added into the existing bulge and the size of new bulge is calculated in the same way as a merger occurred, which will be described later.

\subsubsection{Exchange of Mass, Metal and Angular Momentum}

\begin{figure*}
\centering
\includegraphics[width=\textwidth]{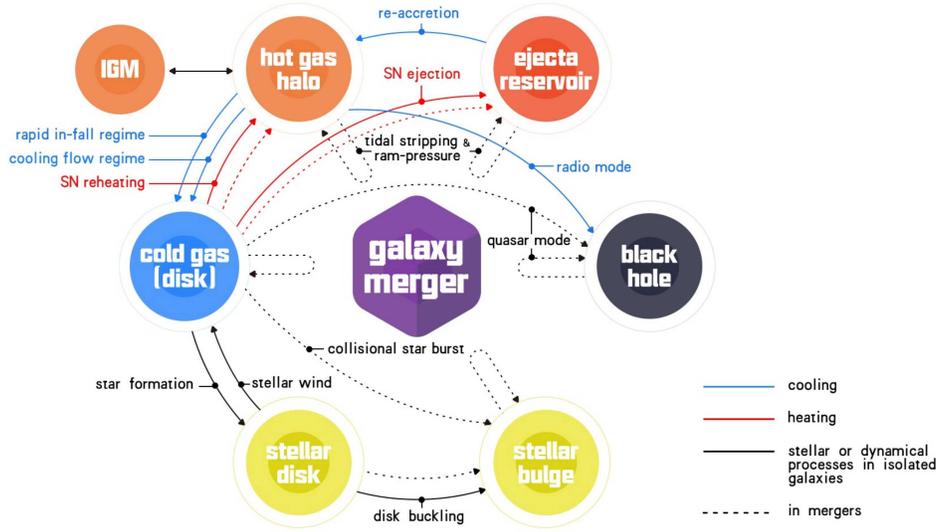}
\caption{A cartoon illustration of the baryonic exchange in {\CODE}. Blue and red lines correspond to the cooling and heating processes in the evolution of galaxy respectively; black lines correspond to stellar or dynamical processes. Solid lines represent physical processes for galaxy evolution in isolation, described in Section \ref{section:isolated}; dashed lines represent physical processes during galaxy interactions, described in Section \ref{section:merger}.}
\label{fig:cycle}
\end{figure*}

When a galaxy evolves in isolation, all physical models should obey conservation laws, including mass and angular momentum conservation. Fig. \ref{fig:cycle} present a cartoon plot illustrating the complex exchange of baryonic matter in our model. Below we summarize in detail the migration of mass, metal, and angular momentum between different phases in our model.

In order to make sure the total mass of galaxy meets a certain fraction $\fb$ of host halo mass, we adjust hot gas mass as
\beqa
\Mhot &= &\fb \Mvir - \Mdiskgas - \Mdiskstar\nonumber \\
& & - \Msphgas - \Msphstar - \Mout - \MBH \nonumber,
\eeqa
and
\beqa
\Mhot \ge 0. \nonumber
\eeqa
Therefore, the exchange of mass can be described by the following equations:
\beqa
\delta \Mhot &= &\Mreturn + \Mreheatdisk + \Mreheatsph \nonumber\\
& &- \Meject - \Mcooleff - \Mradio, \nonumber
\eeqa
\beqa
\delta \Mdiskgas = \Mcooleff - \MSFdisk - \Mreheatdisk, \nonumber
\eeqa
\beqa
\delta \Mdiskstar = \MSFdisk - \Mbar, \nonumber
\eeqa
\beqa
\delta \Msphgas = -\MSFsph - \Mreheatsph, \nonumber
\eeqa
\beqa
\delta \Msphstar = \MSFsph + \Mbar, \nonumber
\eeqa
\beqa
\delta \Mout = \Meject - \Mreturn, \nonumber
\eeqa
\beqa
\delta \MBH = \Mradio. \nonumber
\eeqa

\begin{figure}
\centering
\includegraphics[width=0.5\textwidth]{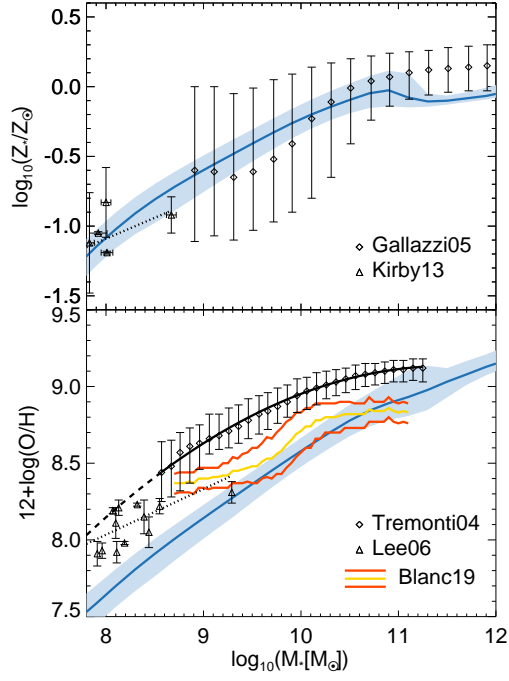}
\caption{Upper panel: the distribution of stellar metallicity in our model galaxies as a function of stellar mass at $z=0$. The blue solid line shows the median values, and the light blue area indicates 16th and 84th percentiles. Diamonds with error bars represent the observational result of \protect\cite{Gallazzi05} from SDSS/DR2; and the triangles with error bars represent the observational result of \protect\cite{Kirby13}, which is derived for local group dwarf galaxies. Lower panel: the distribution of gaseous metallicity in the star-forming galaxies (with specific star formation rate larger than $10^{-11}\yr^{-1}$) as a function of stellar mass at $z=0$. Diamonds with error bars are the observational result of \protect\cite{Tremonti04} from SDSS. Black solid line is their fitting function and dashed line is the extrapolation to lower mass ranges. Triangles with error bars are the observational result of \protect\cite{Lee06} for nearby dwarf galaxies, and the dotted line is their fitting. The yellow and red solid lines are the mean value and $1\sigma$ scatter of the SDSS high mass sample in \protect\cite{Blanc19} respectively.}
\label{fig:metal}
\end{figure}

The ISM of galaxies is enriched by metals released by stellar winds and supernova explosions. Metal enrichment is firstly included into semi-analytic models in \cite{White91}, in which metals are created in stellar evolution and then distributed into other components with mass transfer. The detailed metal exchange in our models is summarized as follows:
\beqa
\delta \Metalhot &= &\Metalreturn + \Metalreheatdisk + \Metalreheatsph \nonumber\\
& &- \Metaleject - \Metalcooleff - \Metalradio, \nonumber
\eeqa
\beqa
\delta \Metaldiskgas &= &\Metalcooleff - \MetalSFdisk - \Metalreheatdisk \nonumber\\
& &+ \yield\MSFdisk, \nonumber
\eeqa
\beqa
\delta \Metaldiskstar = \MetalSFdisk - \Metalbar, \nonumber
\eeqa
\beqa
\delta \Metalsphgas = -\MetalSFsph - \Metalreheatsph + \yield\MSFsph, \nonumber
\eeqa
\beqa
\delta \Metalsphstar = \MetalSFsph + \Metalbar, \nonumber
\eeqa
\beqa
\delta \Metalout = \Metaleject - \Metalreturn, \nonumber
\eeqa
where $\yield = 0.03$ is the so-called ``yield'', controlling the amount of metals created during stellar evolution. In our model, for simplicity, we assume the composition of metal to be solar value, and do not consider different elements independently. Fig. \ref{fig:metal} shows the metallicity in stars and gas as a function of the stellar mass at $z=0$ in our fiducial model. Our model agrees well with observational metallicity for stars. As on gaseous metallicity, our model is marginally in agreement with \cite{Blanc19} for galaxies more massive than $10^{9.5}\Msun$ and is marginally in agreement with the lower limit of \cite{Lee06} for galaxies smaller than $10^{9.5}\Msun$, but way below the observational result of \cite{Tremonti04} for galaxies less massive than $10^{11}\Msun$. This discrepancy between different observational results is caused by systematic uncertainties in the measurement of gaseous metallicity. \cite{Kewley08} demonstrated that depending on the choice of metallicity calibration, the absolute metallicity normalization varies up to 0.7 dex. \cite{Lopez12} and \cite{Blanc19} also showed the discrepancy between different diagnostics, and which one is preferred has not come to a conclusion in the community. As illustrated in the above equations, the amount of gaseous metallicity directly relates to galactic cooling, star formation, supernova feedback and AGN feedback. However, we choose not to use gaseous metallicity as a constrain or calibration for semi-analytic models before we have clearer understanding of the systematic uncertainties in observations.

We follow \cite{Guo11} and also consider the angular momentum transfer among different components. Generally it is assumed that the hot gas share the same spin with its host dark matter halo, and the angular momentum of cooled gas will be transferred into galaxy when stars begin to form. The full exchange of angular momentum could be summarized as the following:
\beqa
\delta \Jdiskgas = \Jcooleff - \JSFdisk - \Jreheatdisk, \nonumber
\eeqa
\beqa
\delta \Jdiskstar = \JSFdisk - \Jbar, \nonumber
\eeqa
\beqa
\delta \Jsphgas = -\JSFsph - \Jreheatsph, \nonumber
\eeqa
\beqa
\delta \Jsphstar = \JSFsph + \Jbar, \nonumber
\eeqa
\beqa
\delta \Jout = \Jeject - \Jreturn. \nonumber
\eeqa

 The disk density profile is often assumed to be exponential, then its scale radius is determined by disk's angular moment. By assuming a flat circular velocity curve,  disk scale radius of a galaxy can be derived with

\beq
\Rdiskgasd = \frac{\Jdiskgas}{2\Vmax\Mdiskgas},
\eeq
\beq
\Rdiskstard = \frac{\Jdiskstar}{2\Vmax\Mdiskstar}.
\eeq

\subsection{Hierarchical Growth}
\label{section:merger}

In the above, we focus on physical processes of the galaxy formation in isolation. In the standard cold dark matter cosmology, structure formation is hierarchical, namely small systems form firstly and then merge to form larger and larger systems. In this section, we will focus on physical processes relevant to mergers.

\subsubsection{Central and Satellite Galaxies}
When two halos approach closer and merge to a single halo in the end, the one with larger mass is called the primary halo, and the smaller one is called substructure of the primary halo and will be eventually tidal disrupted. Following \cite{Springel01}, the galaxy in the primary halo is called central galaxy  (``type 0'' galaxy),  while the galaxy in the substructure is called satellite galaxy. If a satellite galaxy has host subhalo, it is called ``type 1'' galaxy; and if not, it  is called ``type 2'' galaxy. Central galaxies stay at center of the primary halo, while satellite galaxies orbit around their main halo and suffer from tidal stripping as well as ram-pressure stripping.

\subsubsection{Tidal and Ram-Pressure Stripping of Hot Gas}

Tidal radius of a galaxy is a scale beyond which all dark matter and gas get tidally stripped, which can be estimated as

\beq
\Rtidal = \Rvirinfall \frac{\Mvir}{\Mvirinfall},
\eeq
where $\Mvirinfall$ and $\Rvirinfall$ are the virial mass and radius of the satellite's host halo at the infall time\footnote{We define the moment at which the satellite last passes the virial radius of its host central halo as the ``infall'' time.}, respectively.

In addition to the tidal stripping, the hot gas component in satellites suffer ram-pressure stripping due to gas pressure from the primary halo. While ram-pressure stripping acts on both hot and cold gas of satellites, in our model, for simplicity, we only implement this process to more extended hot gas component. The radius, $\Rrp$, that the hot gas of a satellite is stripped by ram-pressure can be estimated by the balance of the pressure from the primary halo and the binding energy of the satellite:
\beq
\rho_{\rm sat} \left( \Rrp \right) V_{\rm vir, sat}^2 = \rho_{\rm cen} \left( R_{\rm orbit} \right) V_{\rm orbit}^2,
\eeq
here $\rho_{\rm sat} \left( \Rrp \right)$ is gas density of the satellite at $\Rrp$, $V_{\rm vir,sat}$ is the virial velocity of the satellite, $\rho_{\rm cen} \left( R_{\rm orbit} \right)$ is the gas density of the central at the position of the satellite, and $V_{\rm orbit}$ is the relative velocity between the satellite and the central. In our model we adopt the minimum of [$\Rtidal$,$\Rrp$] to determine the radius beyond which hot gas of satellite is stripped:
\beq
\Rstrip = min \left( \Rtidal, \Rrp \right).
\eeq

\subsubsection{Dynamical Friction}
When satellites orbit their central galaxies, they are dragged by the surrounding matter due to gravity, gradually lose energy and angular momentum, and will eventually merge with the central galaxy. This process is called dynamical friction (\citealt{Chandrasekhar43}).  As our model is based on subhalo merging trees of N-body simulations, whether a subhalo survives in merger may depend on employed numeric resolution. As assumed by many models that when a subhalo loses its identify in a simulation, namely below 20 particles, it is assumed that the stellar component of the subhalo (i.e. a type 2 galaxy) will still survive for a while because the stellar component is more compact than dark matter. Here we adopt the dynamical friction timescale fitting formula of \cite{Jiang08} (hereafter Jiang08):
\beq
\tfric = \frac{0.94\epsilon^{0.60}+0.60}{2C} \frac{\Mcen}{\Msat} \frac{1}{{\rm ln}\left[ 1+\left( \frac{\Mcen}{\Msat} \right) \right]} \frac{R_{\rm vir,cen}}{V_{\rm c}},
\eeq
where $\epsilon$ is the circularity of the satellite orbit, $C=0.43$ is a constant, $\Mcen$ is the primary dark matter halo mass, $\Msat$ is the satellite mass, including dark matter halo and stars, and $V_{\rm c}$ is the circular velocity. $R_{\rm vir,cen}/V_{\rm c}$ represents free fall timescale in Jiang08, thus we use the virial velocity of central galaxy $V_{\rm vir,cen}$ to estimate $V_{\rm c}$. The circularity $\epsilon$ is calculated through eccentricity $e$ (\citealt{Wetzel10}):
\beq
\epsilon = \sqrt{1-e^2},
\eeq
\beq
e^2 = 1 + \frac{2E_{\rm orbit}J_{\rm orbit}}{\mu \left( G\Mcen\Msat \right)^2},
\eeq
where $E_{\rm orbit}$ and $J_{\rm orbit}$ are the total orbital energy and angular momentum of satellite-central system, $\mu=\Mcen\Msat/(\Mcen+\Msat)$ is the reduced mass.

The dynamical friction timescale is calculated when the satellite first crosses the virial radius of its primary halo. After $\tfric$, this satellite will merge into the central galaxy. This definition of $\tfric$ is consistent with Jiang08.

\subsubsection{Remnants of Mergers}
Numerical simulations demonstrate that the remnants of mergers depend on the orbits and mass ratio of the two merging galaxies (\citealt{Kannan15}). In our model, for simplicity, we follow \cite{White91} to model the remnants of mergers based on the mass ratio of merging galaxies. If the mass ratio $f_{\rm merger}=\Msat/\Mcen$ is greater than $0.3$, it is called ``major merger'', otherwise it is called ''minor merger''. For a major merger, all disk components of both satellite and central are destroyed to form spheroidal components, including cold gas and stars. For a minor merger,  the smaller satellite will be destroyed completely in the end. Its gas component is added into the gaseous disk of the central, and the stellar component is added into the stellar bulge of the central. In both cases, the hot gas  and outflow component will be simply added together respectively. The metals and angular momentum of each component will also be transferred along with mass.

\subsubsection{Black Hole Growth in Mergers}
A very efficient mechanism for the growth of black hole is through mergers, especially major mergers. The violent dynamical environment during mergers is advantaged for black holes to accrete gas. Following \cite{Kauffmann00} and \cite{Croton06}, the growth of central black hole during mergers can be estimated as
\beqa
\delta M_{\rm BH} &= &M_{\rm BH,sat} \nonumber\\
& &+ f \left( \frac{M_{\rm sat}}{M_{\rm cen}} \right) \left[ \frac{M_{\rm cold}}{1 + \left( 280\kms/\Vvir \right) ^2} \right],
\label{eq:Mquasar}
\eeqa
where $M_{\rm BH,sat}$ is the black hole mass of the satellite; $M_{\rm sat}$ and $M_{\rm cen}$ are the masses (cold gas and stars) of the satellite and the central galaxy, respectively; $M_{\rm cold}$ is the total mass of cold gas of two merging galaxies; and $\Vvir$ is the virial velocity of the primary halo. $f$ is a free parameter controlling the growth rate of central black hole.  In our fiducial model, we use  $f=0.03$ and can derive  the tight relation between black hole mass $M_{\rm BH}$ and bulge mass $M_{\rm BH}$ very well.

\subsubsection{Star Burst in Mergers}
Violent interactions during mergers can compress gas and trigger star bursts. We adopt the scheme ``collisional starburst'' of \cite{Somerville01}. In the model, the fraction of the gas transferred into stars during a merger is
\beq
\eburst = 0.56 \left( \frac{\Msat}{\Mcen}  \right)^{0.7},
\eeq
where $\Mcen$ and $\Msat$ are the total mass (cold gas and stars) of the central and satellite galaxy, respectively. Along with the burst, supernova feedback releases huge amount of energy. In a major merger, almost all the remaining cold gas is heated into the hot halo, strongly suppressing further star formation.

\subsubsection{Bulge Formation}
Except material from bar transfer, galaxy merger is another important channel for bulge formation. We use the energy conservation scheme of \cite{Guo11} to estimate the radius of newly formed spheroidal remnant:
\beq
C\frac{GM_{\rm new,bulge}^2}{R_{\rm new,bulge}} = C \frac{GM_{\rm 1}^2}{R_{\rm 1}} + C \frac{GM_{\rm 2}^2}{R_{\rm 2}} + \alpha_{\rm inter} \frac{GM_{\rm 1}M_{\rm 2}}{R_{\rm 1} + R_{\rm 2}},
\label{eq:bulge}
\eeq
where $C=0.5$ is a structure parameter describing the binding energy of each system; $\alpha_{\rm inter}=0.5$ is a parameter describing the interactive energy between systems; $M_{\rm 1}$, $M_{\rm 2}$, and $M_{\rm new,bulge}$ are the masses of progenitor systems and newly formed bulge, respectively; $R_{\rm 1}$, $R_{\rm 2}$, and $R_{\rm new,bulge}$ are the corresponding half-mass radii of progenitor systems and newly formed bulge, respectively. Equ.(\ref{eq:bulge}) is also used to estimate the radius of remnant after bar formation.

\subsection{Luminosity of Galaxies}
\label{section:luminosity}

In order to compare our model with observations, we need to calculate the photometric properties of our model galaxies with a stellar population synthesis model:

\beq
L_{\rm \lambda} = 10^{-0.4 (A_{\rm \lambda,ISM}+A_{\rm \lambda,BC})} \sum_{i=0}^{N_{\rm SP}}{M_{\rm i} L_{\rm \lambda,i}(Z_{\rm i}, t-t_{\rm form,i})},
\label{eq:Ltot}
\eeq
where $L_{\rm \lambda}$ is the luminosity of a galaxy at wavelength $\lambda$; $L_{\rm \lambda,i}(Z,t)$ is the luminosity of unit solar mass stellar population with metallicity $Z$ and age $t$ at wavelength $\lambda$, which is provided by SPSMs; $N_{\rm SP}$ is the total number of stellar populations of the galaxy, $M_{\rm i}$, $Z_{\rm i}$ and $t_{\rm form,i}$ is the initial mass, metallicity and formation time of the $i$th stellar population, all of which are recorded during the running of {\CODE}; $A_{\rm \lambda,ISM}$ and $A_{\rm \lambda,BC}$ are the reddening due to ISM and birth clouds respectively, which are provided by dust extinction models. 

\subsubsection{Yunnan-II Model: Stellar Population Synthesis Model with Binary Evolution}
\label{section:SPSM}

The SPSMs adopted in existing semi-analytic galaxy formation models are mostly based on single star stellar evolution models, e.g. BC03, BC07 with TP-AGBs, and M05. However more than $\sim 50\%$ field stars in our Milky Way are in binary systems. Interactions between stars could produce some specific objects, such as blue stragglers (\citealt{Pols94}) and subdwarf B stars (sdBs, \citealt{Han02}), both of which would alter the Spectral Energy Distribution (SED) of the entire stellar population significantly. In our model, we introduce binary interactions into galaxy formation and evolution by using the Yunnan-II stellar population synthesis model \citep{Zhang04,Zhang05,Zhang10}\footnote{http://www1.ynao.ac.cn/$\sim$zhangfh/}.

Yunnan SPSMs are developed by the Group of Binary Population Synthesis, Yunnan Observatories, Chinese Academy of Sciences. The main differences among different versions of Yunnan models include the stellar evolution models, the evolutionary population synthesis construction method and the condition of binary interactions. The Yunnan-I model (\citealt{Zhang02}) was constructed for stellar populations without binary interactions by using the rapid single stellar evolution algorithms (\citealt{Hurley00}) and used the ``isochrone synthesis'' technique. The Yunnan-II model was constructed for both stellar populations with and without binary interactions by using Monte Carlo simulation and the rapid single and binary stellar evolution algorithms \citep{Hurley00,Hurley02}. In their latest version of Yunnan-II model (\citealt{Zhang10}), the evolutionary population synthesis models of \cite{Han07} which included sdBs were also combined. The Yunnan-III model (\citealt{Zhang13}) was constructed by using single stars' evolutionary tracks obtained from the detailed MESA stellar evolution models (\citealt{Paxton11}). Thus Yunnan-II model has a binary and a single version. The Yunnan-II model with full binary interactions is called {\Ysdb} hereafter and the one which has switched off binary interactions is called {\Yss}. {\Ysdb} has modeled various binary interactions, including mass transfer, mass accretion, common-envelope evolution, collisions, supernova kicks, tidal evolution, and angular momentum loss, etc.

Here we briefly describe the initial conditions and fiducial models adopted in Yunnan-II model and binary stellar evolution (BSE) algorithm, and more details can be found in their papers. In the building of Yunnan-II model, firstly $2.5\times10^7$ sets of initial conditions for binary evolution are created using Monte Carlo method, including the masses of the primary and the secondary star, orbital separation, eccentricity, and metallicities of stars. The initial mass of the primary is generated by assuming a \cite{Chabrier03} IMF with cut-offs of $0.1\Msun$ and $100\Msun$, and a uniform initial secondary over primary mass ratio distribution is adopted. The initial orbital separation is taken to be constant in logarithm and initial eccentricity is distributed in uniform form. In the fiducial model, $50\%$ field stars are assumed to be in binary systems. Actually, if the orbital separation is wide enough, there would be no interactions between component stars and they are effectively single. The stellar evolutionary tracks of generated binaries can be derived though BSE algorithm. Most of binary interactions are modeled through prescription based approach (like galactic evolution models in semi-analytic models), and the properties of produced objects are comparable with many observations (e.g. cataclysmic variables, Algols, double-degenerates, etc, see Section 4.3 of \citealt{Hurley02} for details). Some model parameters have major influence on the binary evolution, such as the efficiency of common-envelope (CE) ejection $\alpha_{\rm CE}$ and Reimers wind mass-loss efficiency $\eta$. $\alpha_{\rm CE}$ describes the efficiency of the orbital energy dissipated by CE and decides whether the outcome of CE evolution is a close binary or a coalescence. The fiducial $\alpha_{\rm CE}$ in Yunnan-II model is 3.0, increased from 1.0, to compensate the lack of internal energy in \cite{Hurley02}. Mass-loss is another crucial process in stellar evolution, and its rate often determines the final state of a binary system. The empirical formula of \cite{Kudritzki78} is used and fiducial efficiency $\eta$ is set to be $0.3$ according to the observations of Galactic globular clusters (\citealt{Iben83}). The influence of input model parameters on the integrated SED of stellar population was discussed in detail in the Section 4 of \cite{Zhang05}. They found that the differences between models with different parameters are less than those produced between models with and without binary interactions.

\begin{figure}
\centering
\includegraphics[width=0.5\textwidth]{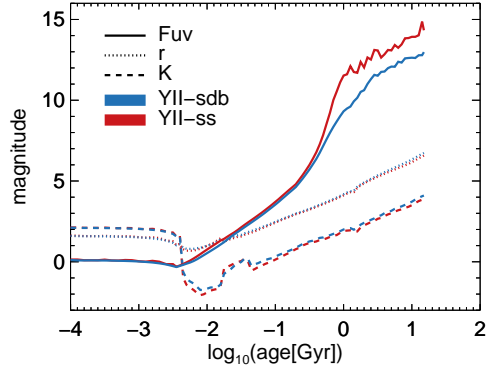}
\caption{Evolution of magnitudes in $\Fuv$, {\sls r}, and K band in {\Ysdb} and {\Yss} assuming $1\Msun$ stellar population with solar metallicity. Different lines and colors distinguish different bands and models as shown in the label.}
\label{fig:yunnan_solar}
\end{figure}

In order to explore the influence of binary evolution on the photometric properties of galaxies, we run our semi-analytic models with and without binary interactions. These two runs share the same semi-analytic models' parameters but adopting different SPSMs, and the comparison between these two runs will be demonstrated in Section \ref{section:photometric}. Fig. \ref{fig:yunnan_solar} shows the magnitudes as a function of age in {\Ysdb} and {\Yss} for $1\Msun$ stellar population with solar metallicity, and three photometric bands are shown: GALEX $\Fuv$ for ultraviolet band, SDSS {\sls r} for optical band, and 2MASS K for infrared band. In {\sls r} and K band, when considering binary interactions, the stellar population is fainter for stars older than $0.01 \Gyr$. This effect is more obvious in K band than in {\sls r} band. {\Ysdb} model is on average $\sim0.13$ magnitude fainter than {\Yss} in K band and $\sim0.07$ magnitude fainter in r band for stars older than $0.01\Gyr$. While in UV band, the situation is reversed,  the stellar population considering binary interactions becomes brighter for stars older than $\sim 1 \Gyr$ ($\sim1.5$ magnitude brighter on average). Because of this difference, binary stars are supposed to be responsible for the UV-excess in elliptical galaxies (\citealt{Han07}). In a short summary, when considering binary interactions in a stellar population synthesis model, luminosity is a bit fainter in the optical ($\sim0.07$ magnitude) and infrared band ($\sim0.13$ magnitude) but is significantly brighter in the far ultraviolet band ($\sim1.5$ magnitude).

\subsubsection{Dust Extinction}
We adopt the dust extinction models of \cite{DeLucia07}, including the extinction by diffuse ISM and birth clouds around young stars. The two reddening terms in Equ.(\ref{eq:Ltot}) are modeled as the following.

The reddening of a slab geometry ISM is given in \cite{Devriendt99}:
\beq
A_{\rm \lambda,ISM} = -2.5\log_{\rm 10}\left[ \frac{1-\exp\left( -\sqrt{1-\wlambda}\tlambda/\cos{i} \right)}{\sqrt{1-\wlambda}\tlambda/\cos{i}} \right],
\eeq
where $\wlambda$ is the albedo of dust (\citealt{Mathis83}); $i$ is the inclination angle and can be estimated according to the direction of spin of a gaseous disk assuming the line-of-sight is along the third axis of N-body simulation; $\tlambda$ is the mean face-on optical depth of the average disk at wavelength $\lambda$:
\beq
\tlambda = \left(\frac{A_{\rm \lambda}}{A_{\rm V}}\right)_{z_{\rm \odot}} \left(\frac{Z_{\rm g}}{Z_{\rm \odot}}\right)^s \left(\frac{\langle N_{\rm H}\rangle }{2.1\times10^{21}{\rm cm^{-2}}}\right) \left( 1+z \right)^{-0.4},
\label{eq:tlambda}
\eeq
\beq
\langle N_{\rm H} \rangle = \frac{\Mdiskgas}{1.4m_{\rm p}\pi\left(aR_{\rm half,disk,gas}\right)}{\rm cm^{-2}}.
\label{eq:nH}
\eeq

The first part of Equ.(\ref{eq:tlambda}) describes the extinction at the given wavelength relative to the extinction at $5500 \AA$ (\citealt{Mathis83}); the second part describes the dependence of extinction on gas metallicity ($s = 1.35$ for $\lambda < 2000 \AA$ and $s = 1.6$ for $\lambda>2000 \AA$, \citealt{Guiderdoni87}); the last part describes the dependence on redshift (\citealt{Guo09}). $\langle N_{\rm H} \rangle$ is the mean H column density, and $a = 1.68$ in Equ.(\ref{eq:nH}).

The reddening for the birth clouds around young stars is given in \cite{Charlot00} and \cite{DeLucia07}:
\beq
A_{\rm \lambda,BC} = -2.5\log_{\rm 10}e^{-\tau_{\rm \lambda,BC}},
\eeq
\beq
\tau_{\rm \lambda,BC} = \tlambda \left( \frac{1}{\mu}-1 \right) \left( \frac{\lambda}{5500 \AA} \right)^{-0.7},
\eeq
where $\mu$ is a random number drawn from a Gaussian distribution with a center 0.3 and a width 0.2, truncated at 0.1 and 1 (\citealt{Kong04}).

\subsection{Model Calibration by the Observations}
\label{section:calibration}

\begin{table*}
\bc
\caption{Free parameters in {\CODE} and their fiducial values.}
\begin{tabular}{ccc}
\hline
Parameter & Description & Fiducial Value\\
\hline
$\alpha$ & Star formation efficiency, Equ.(\ref{eq:dMsfdisk}) & 0.02\\
$\epsilon$ & Amplitude of SN reheating efficiency, Equ.(\ref{eq:ereheat}) & 6.5\\
$\beta_{\rm 1}$ & Slope of SN reheating efficiency, Equ.(\ref{eq:ereheat}) & 3.5\\
$\eta$ & Amplitude of SN ejection efficiency, Equ.(\ref{eq:eeject}) & 0.32\\
$\beta_{\rm 2}$ & Slope of SN ejection efficiency, Equ.(\ref{eq:eeject}) & 3.5\\
$\Vmaxc$ & Characteristic $\Vmax$ in SN efficiency, Equ.(\ref{eq:ereheat}) and Equ.(\ref{eq:eeject}) & 90$\kms$\\
$\gamma$ & Outflow return efficiency, Equ.(\ref{eq:Mreturn}) & 0.3\\
$f$ & Black hole growth efficiency in mergers, Equ.(\ref{eq:Mquasar}) & 0.03\\
$\kappa$ & Quiescent Black hole accretion rate, Equ.(\ref{eq:Mradio}) & $8.64\times 10^{-4}$\\
\hline
\label{tab:parameters}
\end{tabular}
\ec
\end{table*}

In our whole set of physical models, a few parameters are set free and need to be calibrated according to observational data. Similar to \cite{Croton06} and \cite{Guo11}, we have 9 free parameters in our models which are summarized in Tab.(\ref{tab:parameters}). We adjust these parameters within the plausible ranges provided by \cite{Croton06} to fit the stellar mass function at $z=0$, and simultaneously to obtain a reasonable $M_{\rm sph,*}-\MBH$ scaling relation and stellar disk sizes.  The fiducial values of these parameters are listed in Tab.(\ref{tab:parameters}). By adopting these fiducial values and {\Ysdb}, we build our fiducial model.

\begin{figure}
\centering
\includegraphics[width=0.5\textwidth]{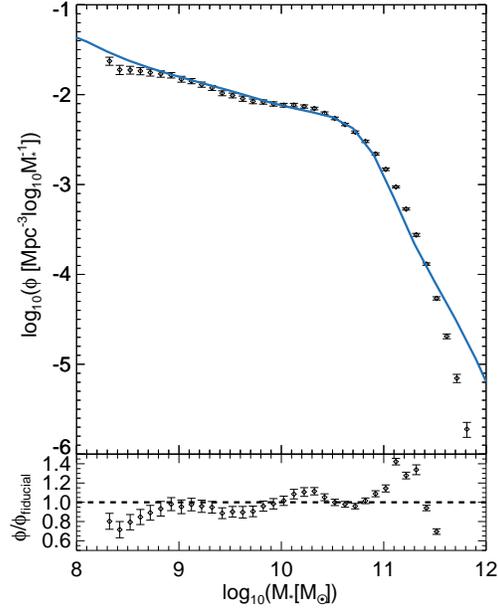}
\caption{Galaxy stellar mass functions at $z=0$. The blue line shows the result of our fiducial model, and the diamonds with error bars show observational result of \protect\cite{Li09}, using SDSS/DR7 data. The lower panel show ratios between ours and observational results. The black horizontal dashed line indicates unity.}
\label{fig:smf_parameter}
\end{figure}

\begin{figure}
\centering
\includegraphics[width=0.5\textwidth]{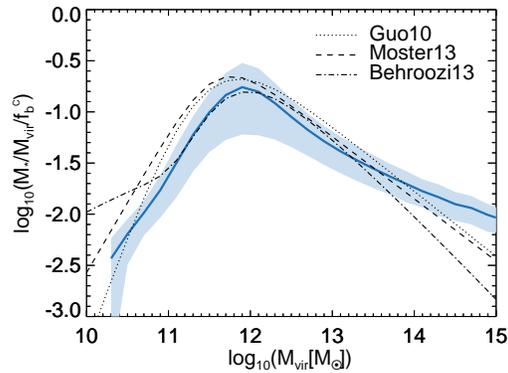}
\caption{Relation between central galaxy stellar mass and the host halo mass. The blue solid line shows the mean value in each mass bin for central galaxies in our fiducial model. The light blue shadow indicates 16th and 84th percentiles. Black lines show results from subhalo abundance matching models  (dots, \protect\citealt{Guo10}; dashed, \protect\citealt{Moster13}; dot-dashed, \protect\citealt{Behroozi13}).}
\label{fig:MstarMhalo}
\end{figure}

Fig. \ref{fig:smf_parameter} presents galaxy stellar mass function  at $z=0$ in our fiducial model. We can see that the predicted stellar mass function agrees very well with the observational result of \cite{Li09}. Only at the most massive end, with $M > 10^{11.5}\Msun$, the number density is slightly higher than observation. To have a better understanding of the inconsistency, we plot the stellar mass of central galaxies against their host halo mass in Fig. \ref{fig:MstarMhalo}. Compared to subhalo abundance matching results of \cite{Guo10}, \cite{Moster13} and \cite{Behroozi13}, our model predicts too massive galaxies in halos $> 10^{13.5} \Msun$. Tuning parameters in the reasonable range can only improve the situation.  Note, such an excess at most massive end also exists in other semi-analytic models, such as \cite{Croton06} and \cite{Guo11}. However more recent semi-analytic models seem overcoming this problem with improved AGN feedback models (\citealt{Croton16} and \citealt{Henriques15}). We will explore this in our future work. 
 
\begin{figure}
\centering
\includegraphics[width=0.5\textwidth]{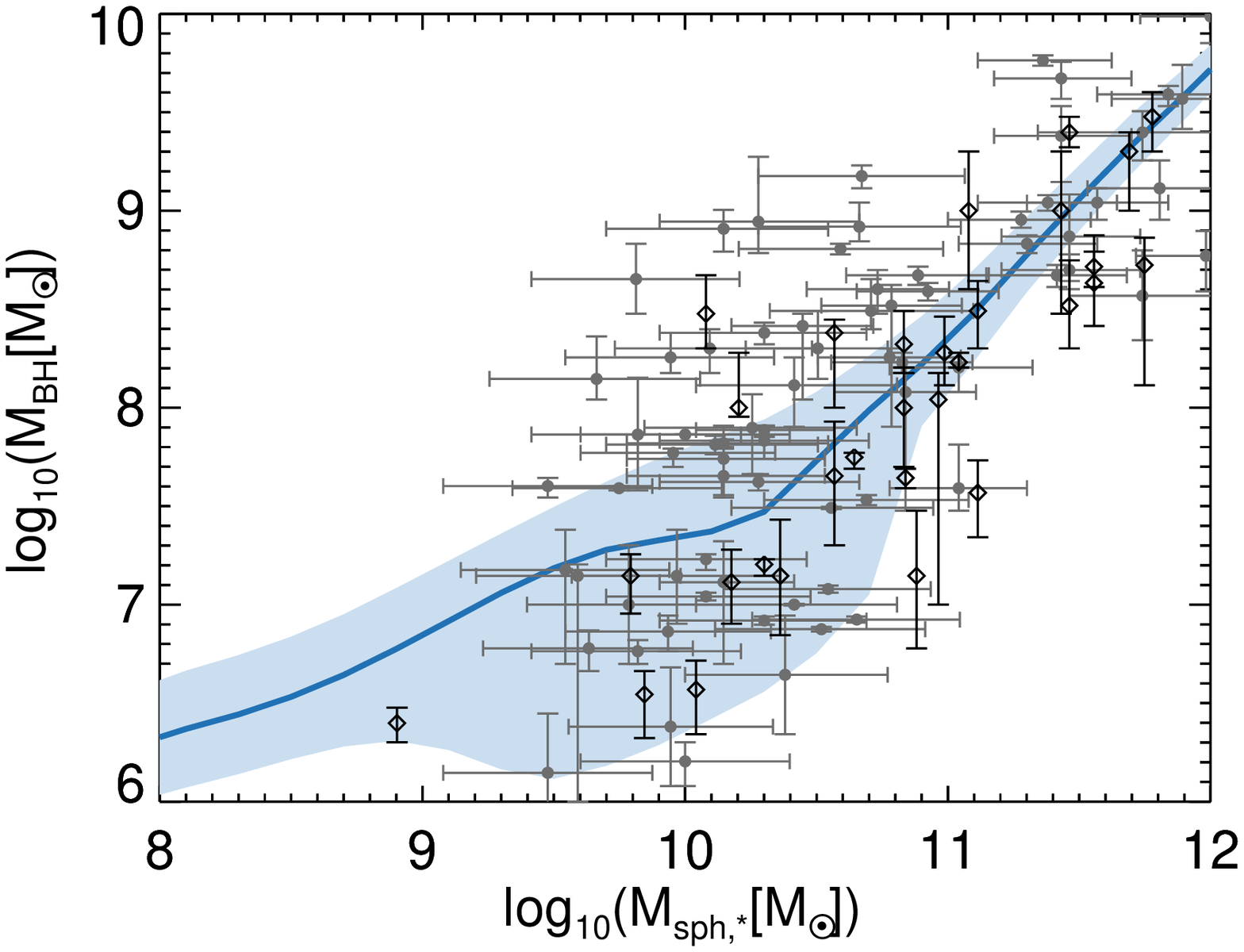}
\caption{The relation between black hole and stellar bulge masses at $z=0$. The blue line shows median values of our fiducial model, and the light blue shadow indicates 16th and 84th percentiles. The black diamonds with error bars are the observational result of \protect\cite{Haring04}. The gray filled cycles with error bars show the observational result of \protect\cite{Scott13}.}
\label{fig:MBHMbulge}
\end{figure}

Fig. \ref{fig:MBHMbulge} shows the black hole to stellar bulge mass scaling relation in our fiducial model. Our results agree well with the observation yet with slightly smaller scatter than observations. This is acceptable given the fact that the systematic errors of the observation is large. 
 
\begin{figure}
\centering
\includegraphics[width=0.5\textwidth]{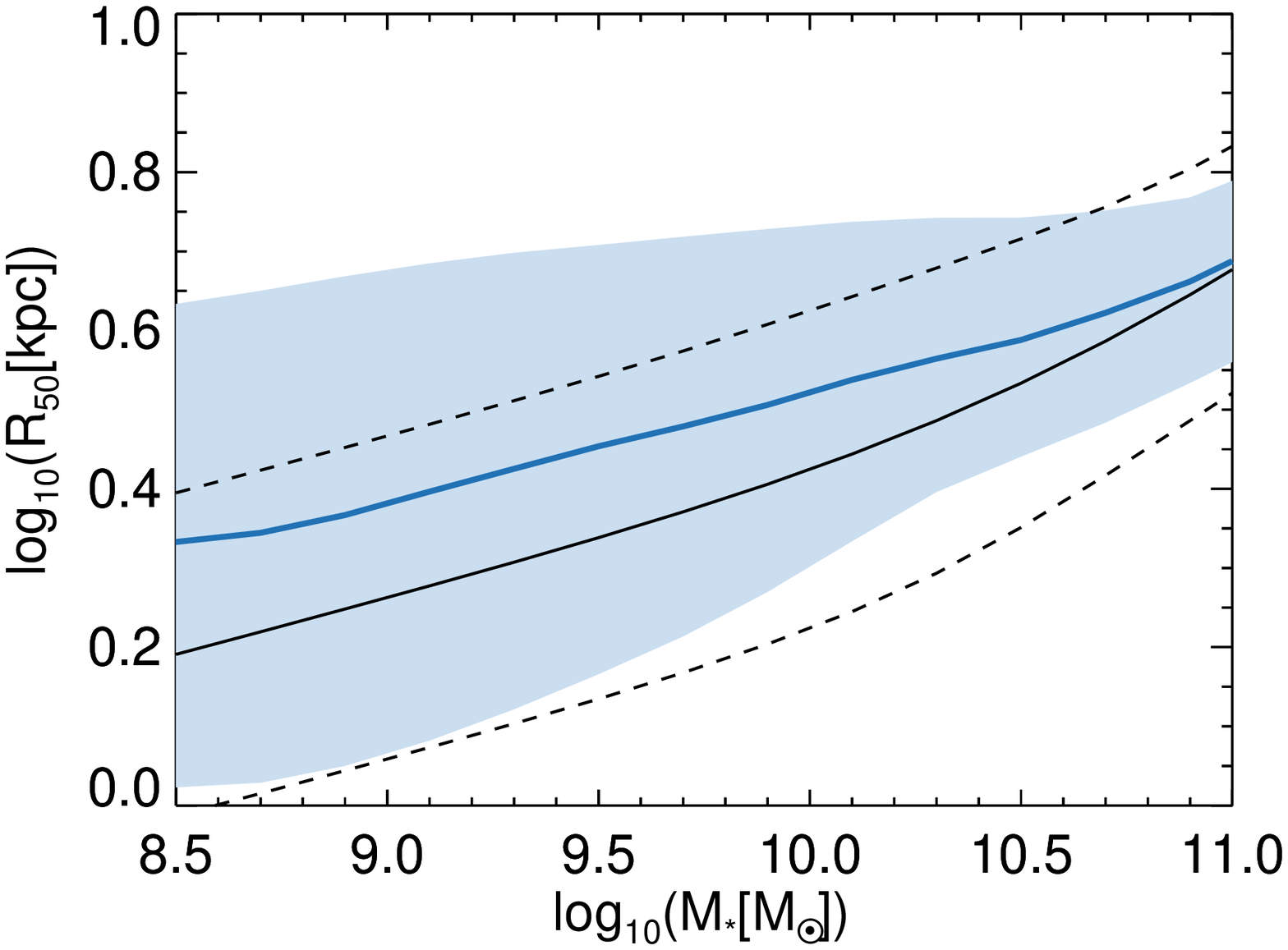}
\caption{The distribution of the half mass radius of stellar disks for late type galaxies ($\Mdiskstar/(\Mdiskstar+\Msphstar) > 0.8$) as a function of stellar mass in our fiducial model. The blue solid line shows the median values of our model galaxies, and the light blue shadow indicates 16th and 84th percentiles. The black solid line shows the fitting formula of the observational results from \protect\cite{Shen03}, and the dashed lines show $1\sigma$ scatters.}
\label{fig:Rdisk}
\end{figure}

Fig. \ref{fig:Rdisk} is the distribution of the half mass radius of stellar disks for late type galaxies in our fiducial model. Our model prediction is shown as the blue solid curve. While the observational data from \cite{Shen03} is presented as the black solid and dashed lines. Again, the radius of stellar disks in our model agree well with the observational data in most mass range. There is a slightly disagreement at small mass end, this is acceptable given the fact that the uncertainty in measurement of galaxies size is also large at these mass scales.  

\section{Results}
\label{section:result}

\subsection{The Influence of Binary Evolution}
\label{section:photometric}

In the following, we examine the influence of adopting Yunnan-II SPSM on the galaxy luminosity function at different bands and galaxy color.

\subsubsection{Luminosity Functions}

\begin{figure*}
\centering
\includegraphics[width=1.0\textwidth]{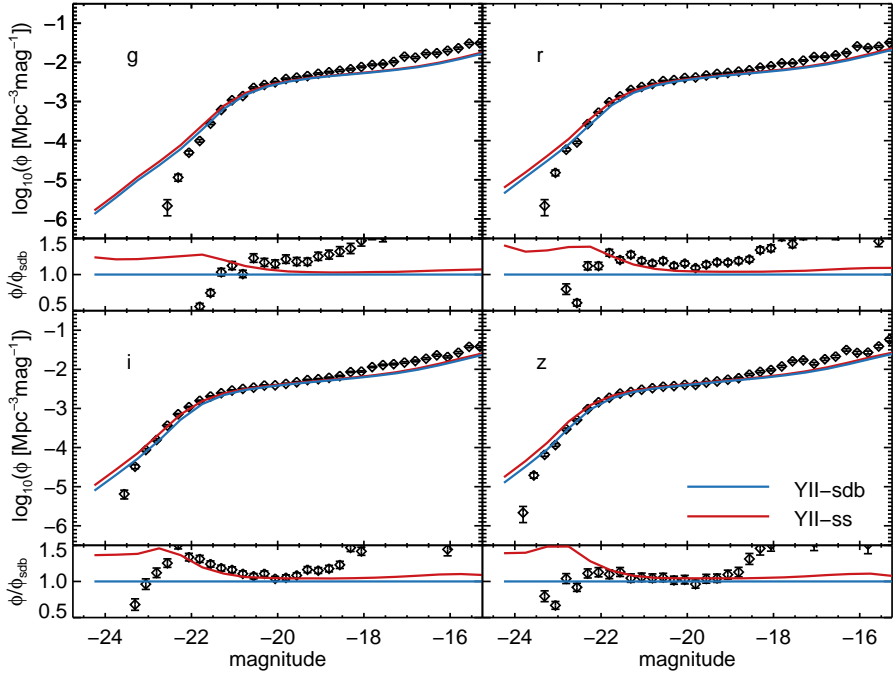}
\caption{The luminosity function of galaxies at $z=0$ in the SDSS {\sls g}, {\sls r}, {\sls i}, and {\sls z} bands. The red and blue solid lines show results with Yunnan-II SPSMs {\Ysdb} and {\Yss}) respectivly. The diamonds with error bars show the observational result from SDSS/DR2 (\protect\citealt{Blanton05}). In each panel, the ratio between two Yunnan-II models and the ratio between SDSS and {\Ysdb} results are showed at the bottom.}
\label{fig:LF}
\end{figure*}

In Fig. \ref{fig:LF}, we present the luminosity function of galaxies at $z=0$ in the SDSS {\sls g}, {\sls r}, {\sls i}, and {\sls z} bands. The results with {\Ysdb} and {\Yss} are shown as blue and red lines respectively, while the observed data from SDSS are shown as diamonds with error bars. The ratio of the observed data or the {\Yss} over {\Ysdb} model are shown in bottom panels. We see that the luminosity functions in both models agree well with the observational result over most magnitude scales. At the brightest end, the model prediction is higher than the observational data. Cosmic variance could be partly responsible for that. At the faint end with $M >-18$, our model prediction is lower than the observations because of the resolution is not enough for these faint galaxies in the Milleniuum simulation. Furthermore, in all bands, the difference between two Yunnan-II models are very tiny at faint galaxy end (less than $10\%$), but the difference can be $20\% \sim 50\%$ for galaxies above the character luminosity (the ``knee'' of luminosity function). Binary interactions have the similar behavior for galaxies' luminosity and simple stellar population (as shown in Fig. \ref{fig:yunnan_solar}). They make galaxies fainter in all optical bands, hence lower the luminosity function. Besides, this effect gets a little bit more obvious at longer wavelength. After considering binary interactions, the average shifts of luminosity functions along x-axis are $0.10$, $0.14$, $0.15$ and $0.16$ magnitude for {\sls g}, {\sls r}, {\sls i}, and {\sls z} bands respectively.

\begin{figure}
\centering
\includegraphics[width=0.5\textwidth]{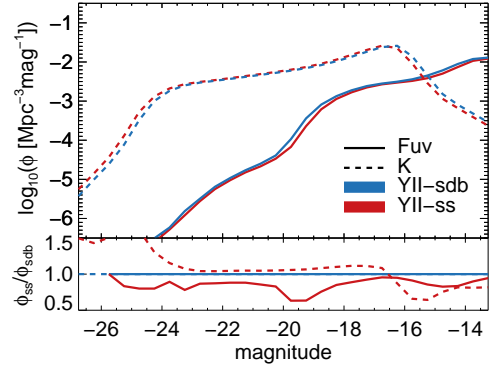}
\caption{In the top panel, The luminosity function of galaxies in our fidiucial model at $z=0$ in the GALEX $\Fuv$ band and 2MASS K band are presented as solid and dashed lines individually. The two Yunnan-II SPSMs models, {\Ysdb} and {\Yss} are indicated as blue and red respectively. The ratio of the results from two SPSMs models are presented in the bottom panel. }
\label{fig:LF_Fuv}
\end{figure}

As shown in Fig. \ref{fig:yunnan_solar}, the evolution of a simple stellar population with two Yunnan-II models only differ with each other by $\sim0.07$ magnitude in optical bands, which is agree with what we see in  Fig. \ref{fig:LF}. Also, it is expected that the difference would be larger in K band and $\Fuv$ band, especially when the population gets older. Therefore, we checked that in Fig. \ref{fig:LF_Fuv}, in which we presented the luminosity function of galaxies from two SPSMs models in GALEX $\Fuv$ band and 2MASS K band with solid and dashed lines, respectively. The ratio of two models are also presented in the bottom panel. In $\Fuv$ band, the results of binary evolution model {\Ysdb} is always higher than {\Yss}, this indicates galaxies get brighter if the binary evolution is included. The situation is reversed in $K$ band. Galaxies with binary evolution gets fainter, as what we see in the optical bands, but more obvious. After considering binary interactions, the average shifts of luminosity functions along x-axis are $-0.20$ and $0.18$ magnitude for $\Fuv$ and K bands respectively.

\begin{figure}
\centering
\includegraphics[width=0.5\textwidth]{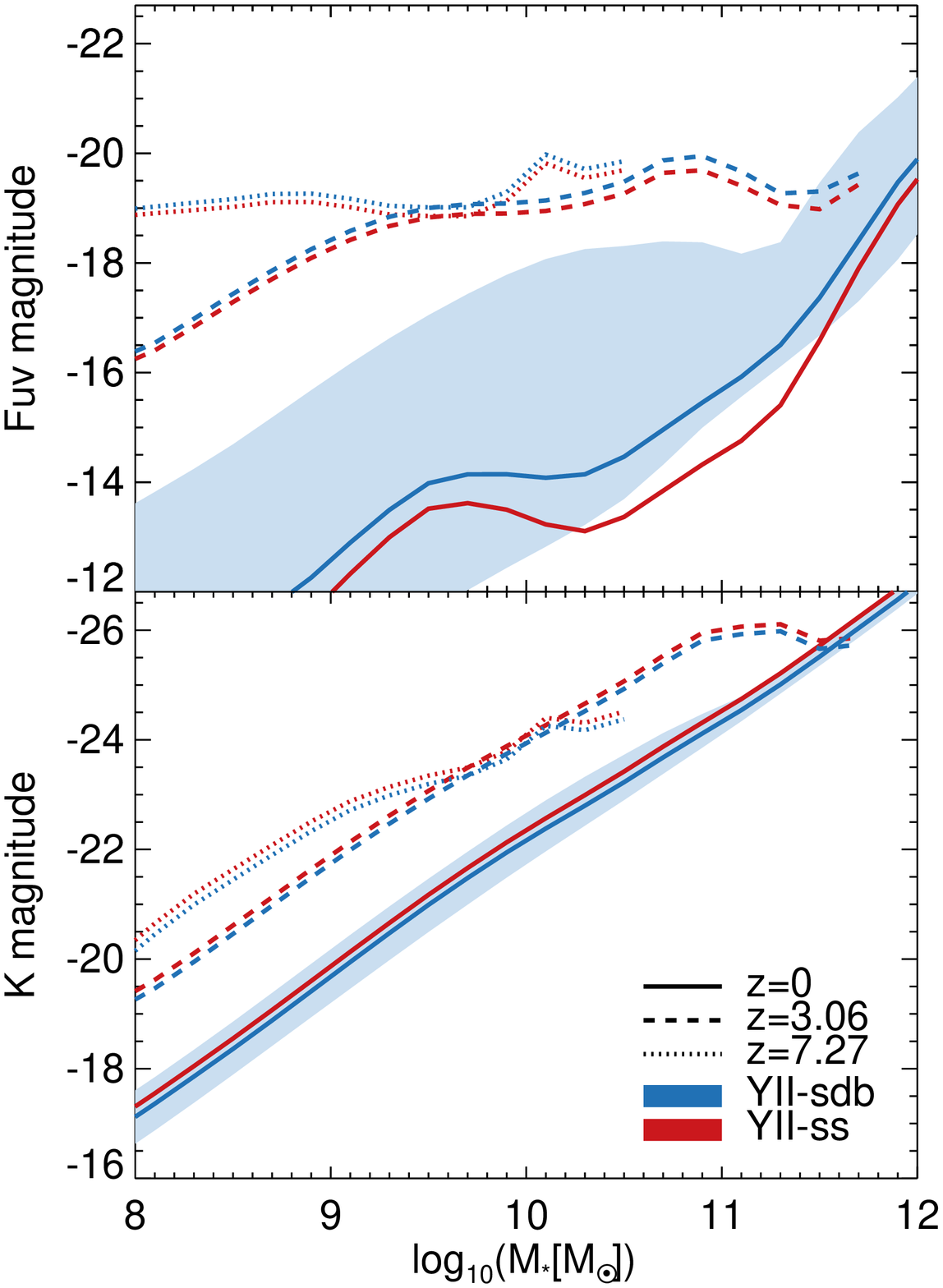}
\caption{ The Luminosity  of our model galaxies in $\Fuv$ and K band are presented as a function of the stellar mass in top and bottom panels respectively. The median values in each mass bin from two Yunnan-II models,{\Ysdb} and {\Yss}) are showed as blue and red color. The results from different redshifts $z=0, 3.06, 7.27$ are indicated with solid, dashed and dotted curves individually. The blue shaded regions indicate 16th and 84th percentiles for the {\Ysdb} model at $z=0$.}
\label{fig:MstarL}
\end{figure}

To further understand the difference on the luminosity function between two Yunnan-II models above, we explore the relation between galaxies' magnitude and their stellar mass in $\Fuv$ and K band for two Yunnan-II models in the upper and bottom panel of Fig. \ref{fig:MstarL} respectively. It is notable that, for a given stellar mass galaxy, the $\Fuv$ luminosity is always brighter when the binary evolution is considered. Specially, for galaxies with $\sim 10^{10.5}\Msun$ stellar mass at $z=0$, the difference is significant and reaches $\sim1.5$ magnitude. This overall excess origins from binary interactions, which will bring about an excess in $\Fuv$ band in old stellar populations. We have check that most of this excess is contributed by early type galaxies, and the $\Fuv$ luminosity of late type galaxies is less influenced by binary interactions. The results at redshift $z=0, 3.06, 7.27$ are also shown by different type of lines. It is obvious that the difference is smaller at higher redshift. This is because the difference between two SPSMs gets larger when the populations in galaxies get older. The results is reversed for K band in the bottom panel. The K magnitude in {\Yss} is always brighter than that with binary considered for all mass scales and redshifts. The suppression of the formation of red giants by the interactions in binary is supposed to be responsible for that.

In summary, the inclusion of binary evolution will suppress the galaxies' luminosity in optical and infrared bands (SDSS bands and 2MASS K band) slightly, but increase the luminosity at the ultraviolet band significantly. The displacements of luminosity function along magnitude are $0.10\sim0.16$, $0.18$ and $-0.20$ in SDSS bands, K band and $\Fuv$ band respectively. However, given the big scatter in the models (blue shaded region in Fig. \ref{fig:MstarL}) and huge uncertainties in the observation, the difference between two SPSMs models are still not distinguishable in the current observational data.

\subsubsection{Color Distribution}

\begin{figure*}
\centering
\includegraphics[width=1.0\textwidth]{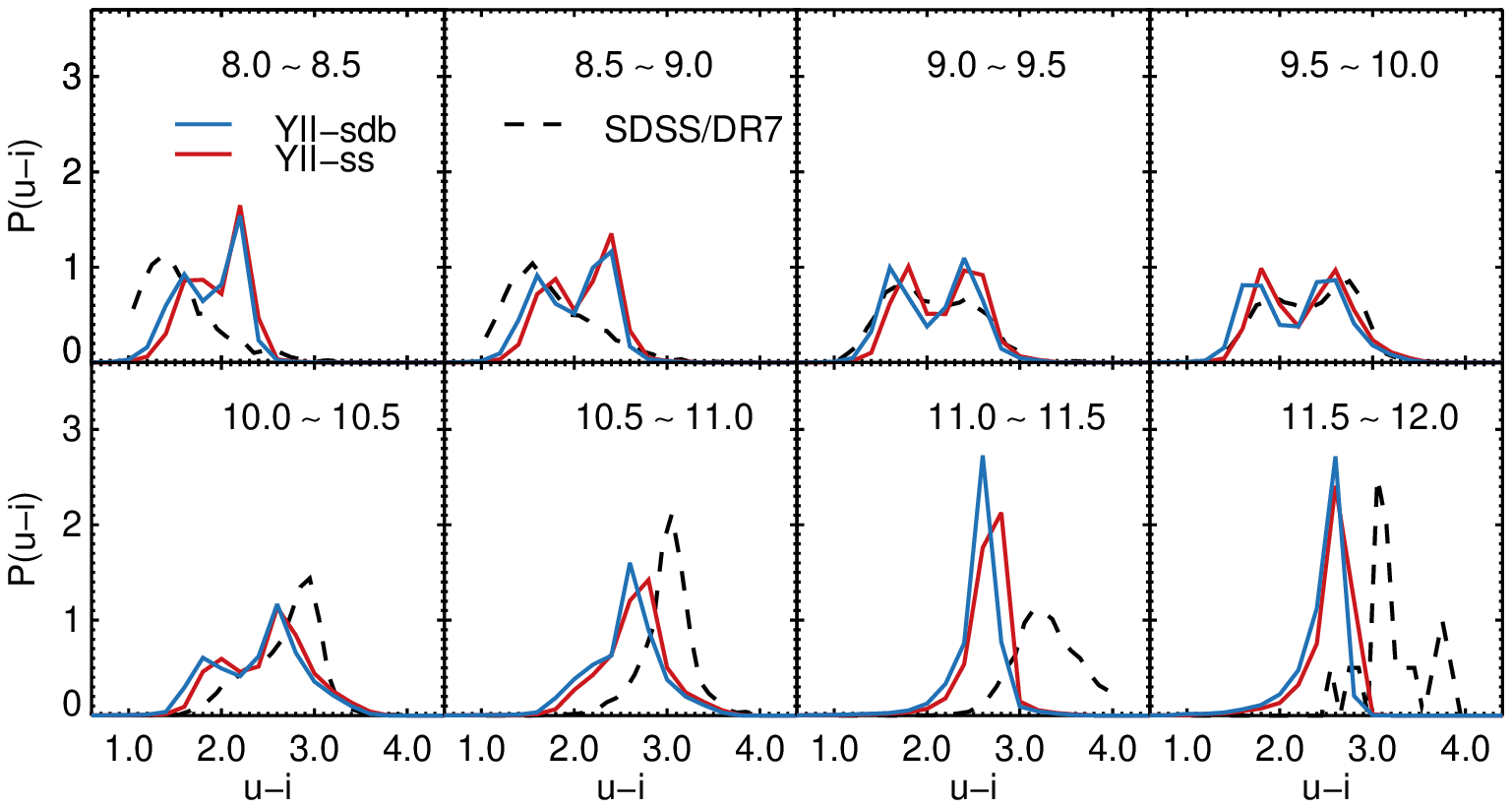}
\caption{{\ui} color distribution of our model galaxies in different stellar mass ranges. The results of Yunnan-II models ,{\Ysdb} and {\Yss}, are presented as red and blues.}
\label{fig:colordis}
\end{figure*}

Furthermore, we checked the color index {\ui} distribution of our model galaxies in different stellar mass bins in Fig. \ref{fig:colordis}.  The results from two SPSMs models, {\Ysdb} and {\Yss}, are presented with blue and red solid curves respectively; and the observational data from SDSS is shown as black dashed curves. Generally, our fiducial model with {\Ysdb} can recover the bimodality of color distribution in SDSS data in stellar mass range $[10^{9.0},10^{10.5}]\Msun$. But at the massive end, our model predictions are bluer than observations, and at the faint end, our model galaxies are redder than observed result. As other semi-analytic model pointed out, to get the correct prediction on the color distribution, some physical models need to be updated or even some new physics are needed (\citealt{Henriques15}). Moreover, galaxies in {\Ysdb} is $\sim0.1$ bluer.

\begin{figure*}
\centering
\includegraphics[width=1.0\textwidth]{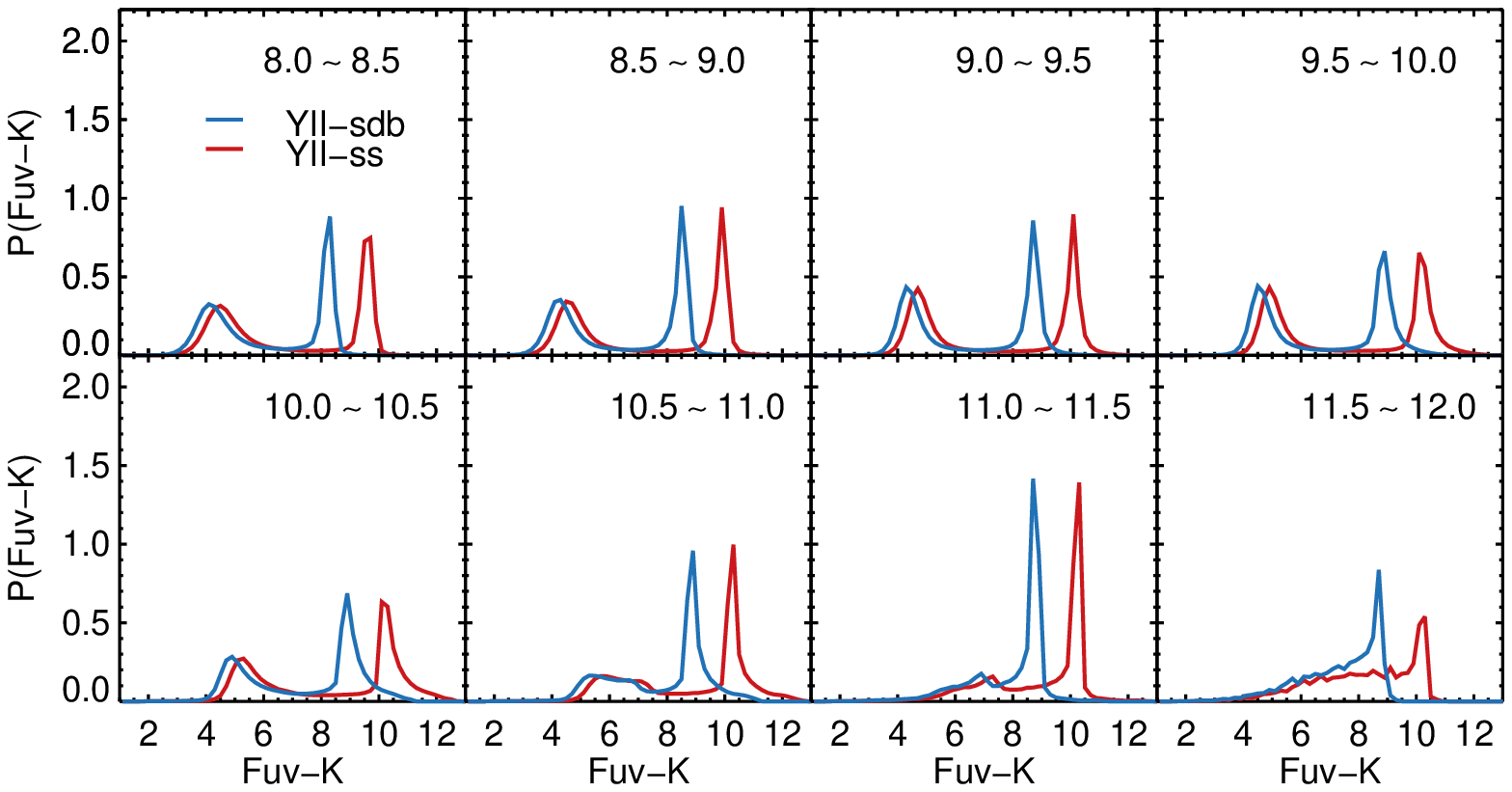}
\caption{{\Fuvk} color distribution of our model galaxies in different stellar mass ranges. The results of Yunnan-II models ,{\Ysdb} and {\Yss}, are presented as red and blues.}
\label{fig:colordis_Fuv}
\end{figure*}

As the influences of binary evolution on galaxies' flux in $\Fuv$ and K band is bigger, it is supposed that the color index between these two bands will change  significantly when the binary evolution is included. In Fig. \ref{fig:colordis_Fuv}, the color index $\Fuv-$K is presented for both Yunnan-II models. The bimodality feature is apparent. With binary evolution, the blue peak is shifted towards blue end by $\sim0.4$, and the red peak is shifted towards blue end by $\sim 1.6$, which is three times larger than blue peak. The galaxies in {\Ysdb} is brighter in $\Fuv$ band because of the existence of hot stars, so they have smaller $\Fuv-K$ and shifted to the left in every panels comparing to the model prediction of {\Yss}. The red peak corresponds to old galaxies which are influenced substantially by binary evolution in $\Fuv$ band (as shown in Fig. \ref{fig:yunnan_solar}), thus the shift of red peak is much larger than the blue peak. If the observational $\Fuv$ magnitudes could be constrained better in the future, it is very promising that we could tell the difference between these two model predictions with the observations.

\subsection{Remnants of Binary Evolution}

\begin{figure}
\centering
\includegraphics[width=0.5\textwidth]{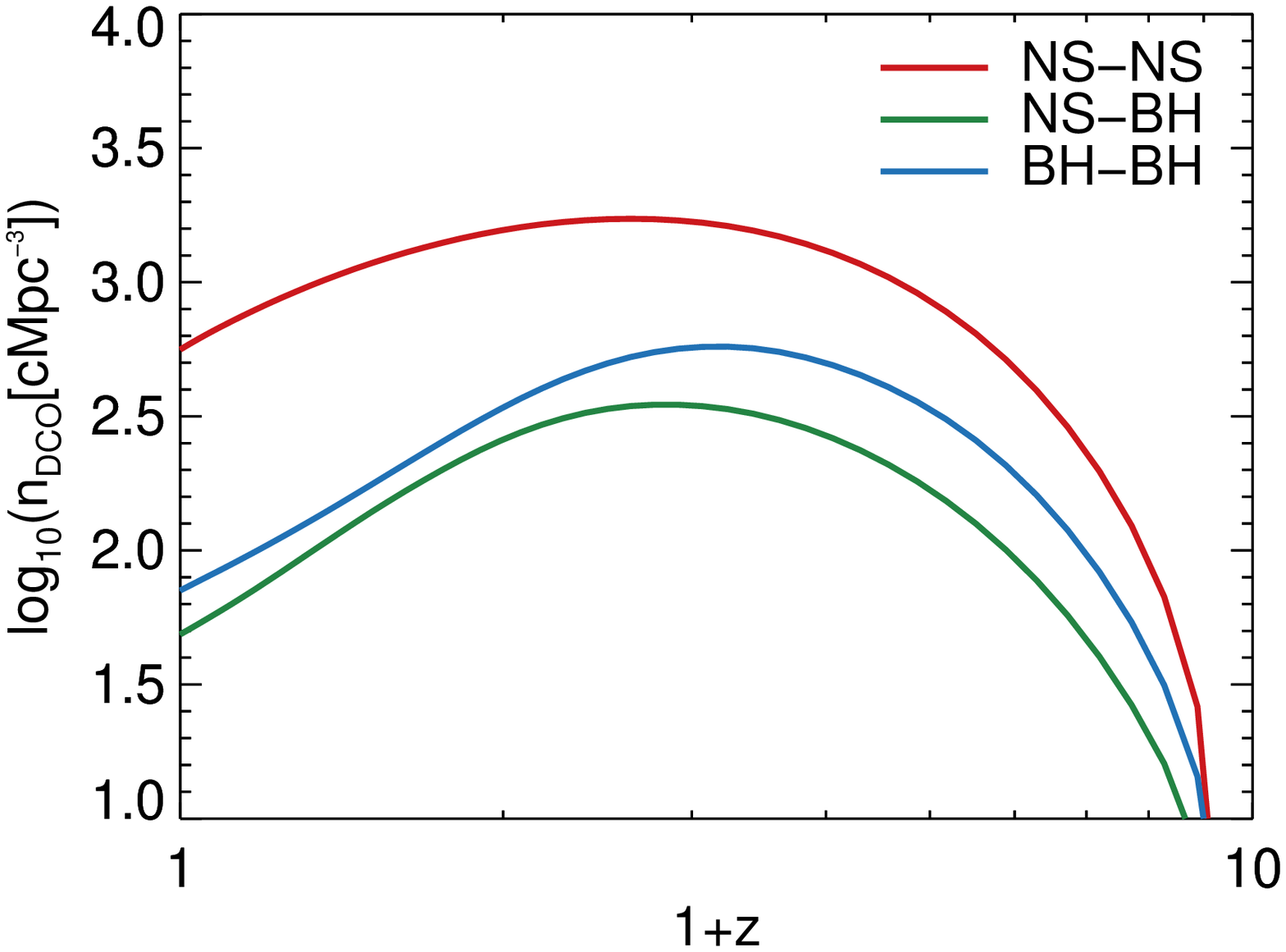}
\caption{The predicted cosmic number per cubic comoving volume of double compact objects (DCOs, including NS-NS, NS-BH and BH-BH) as a function of redshift in our fiducial model. Different colors show results for different DCOs as indicated in the label.}
\label{fig:NNz}
\end{figure}

The remnants of the evolution of binary stars (e.g. NS-NS, NS-BH, BH-BH) are thought to be associated with many high energy events, like type Ia supernovae, short gamma-ray bursts, kilnovae and gravitational wave events. In our model, using Yunnan-II binary SPSM, we are able to study these remnants in a cosmological context. As an example, in Fig. \ref{fig:NNz} we show the cosmic number density of the double compact objects (hereafter DCOs), i.e. NS-NS, NS-BH and BH-BH as a function of cosmic time. With the number density and merger rate of these DCOs, we could study some high energy events associated with mergers of them, e.g. gravitational wave signals, kilonovae, short gamma-ray bursts etc. We will present these results in our forthcoming papers.

\section{Conclusions}
\label{section:conclusion}

In this work, we develop a new semi-analytic galaxy formation model GABE. The model is based on merger trees provided by cosmological N-body simulations. It succeeds many physical recipes from existing successful models and also includes some new ingredients. For example, it adopts Yunnan-II SPSM which carefully considers binary evolution, and a cooling table which takes into account UV heating properly. 

When calibrating our model to the observed stellar mass function, our model reproduces a large body of observations. We explore the influence of Yunnan SPSM model with binary evolution on the galaxy luminosity function and color. Our finding is that, with binary SPSM, the influence on luminosity is quite small over different SDSS bands, and is more substantial in infrared band and ultraviolet band. The shifts of luminosity function along magnitude after considering binary evolution are $0.10\sim0.16$, $0.18$ and $-0.20$ magnitude in SDSS bands, K band and $\Fuv$ band respectively. Galaxies appear to be bluer with binary evolution, especially when $\Fuv$ band is under consideration. The shift of {\ui} color index is about $-0.1$. The shift of the blue peak of $\Fuv-K$ color index is about $-0.4$, and the one of the red peak is about $-1.6$, three times larger than the blue peak.

This is the first paper of our series works. We have used {\CODE} to simulate the gravitational wave signals of supermassive black holes in \cite{Wang19}. Besides, the new feature of our model allows us to predict the population of double compact objects (e.g. NS-NS, NS-BH, BH-BH) which are expected to be associated with some high energy events in a cosmological context. For example, the short gamma-ray bursts which are expected to be products of NS-NS or NS-BH mergers; gravitational wave events which are supposed to be generated during mergers of two compact objects in binary systems. In addition, we will be able to explore the relation between these high energy events and their host galaxies. We will present these studies in our future works.

\normalem
\begin{acknowledgements}

We acknowledge Zhanwen Han for reading our draft and providing some very useful comments and suggestions. We also would like to thank the anonymous referee for an insightful report which led to improvements in the presentation and content of our paper. We acknowledge support from the National Key Program for Science and Technology Research and Development (2015CB857005, 2017YFB0203300) and the Chinese Natural Science Foundation (NSFC) grants (11390372, 11425312, 11503032, 11573031, 11851301, 11873051, 11573062, 11521303, 11390734, 11573033, 11622325 and 11573030). Besides, FZ acknowledges support from the YIPACAS Foundation (grant 2012048) and the Yunnan Foundation (grant 2011CI053). QG is  supported by the Newton Advanced Fellowship. JP acknowledges support from the National Basic Research Program of China (program 973 under grant no. 2015CB857001).

\end{acknowledgements}
  
\bibliographystyle{raa}
\bibliography{gabe}

\end{document}